\documentclass[showpacs,preprintnumbers,amsmath,amssymb,twocolumn]{revtex4-1}
\usepackage{graphicx}
\usepackage{tabularx}
\usepackage{booktabs}
\usepackage{mathtools}
\usepackage{enumerate}
\usepackage{epstopdf}
\usepackage[colorlinks=true,allcolors=blue]{hyperref}
\usepackage[all]{hypcap}
\usepackage{float}
\usepackage[usenames,dvipsnames]{color}
\usepackage{diagbox}

\begin{document}
   \title{New Approach to Evolving Gravitational Waves in Loop Quantum Cosmology}

   \author{Wen-Hsuan Lucky Chang}
   \email{f00222018@ntu.edu.tw}
   
   \author{Jiun-Huei Proty Wu}
   \email{jhpw@phys.ntu.edu.tw}

   \affiliation{Department of Physics, Institute of Astrophysics, and Center for Theoretical Physics,\\
					National Taiwan University,	No.1 Sec.4 Roosevelt Road, Taipei 10617, Taiwan}

%	\date{\today}

\begin{abstract}
	With the observational advance in recent years, primordial gravitational waves (GWs), known as the tensor-mode cosmic perturbations, in the Loop Quantum Cosmology (LQC) are becoming testable and thus require better framework through which to bridge between the observations and the theories. In this work we present a new formalism that employs the transfer functions to bring the GWs from any epoch, even before the quantum bounce, to a later time, including the present. The evolutionary epochs considered here include the possible deflation, quantum bounce, and inflation. This formalism enables us to predict more accurately the GW power spectrum today. With the ADM formalism for the LQC background dynamics, our approach is equivalent to the commonly used Bogoliubov transformations for evolving the primordial GWs, but more transparent for discussions and easier to calculate due to its nature of being linear algebra dealing with linear perturbations. We utilize this advantage to have resolved the IR suppression problem. We also propose the field-free approximation for the effective mass in the quantum bounce epoch to largely improve the accuracy in the predicted GW power spectrum. Our transfer-function formalism is general in dealing with any linear problems, and thus expected to be equally useful under other context with linearity.
\end{abstract}

\maketitle

% -----------------------------------------------------------------------
\section{Introduction}\label{sec01}

Loop Quantum Cosmology (LQC) is a simplified theory of Loop Quantum Gravity (LQG) with the cosmological principle: the universe is homogeneous and isotropic on large scales \cite{Bojowald2002a}. It is based on the Friedmann-Robertson-Walker (FRW) model with the quantum corrections. With the addition of a scalar field, the singularity problem in the standard cosmology \cite{Hawking1970} can be avoided by a quantum bounce \cite{Bojowald2001a}. Therefore LQC provides a clean link between the present universe and the `parent universe' \cite{Ashtekar2006b,Ashtekar2006c,Ashtekar2006d}. 
In this study, we presume the curvature constant in the FRW model to be zero because the observational results shows that the spacial geometry of our universe is consistent with being flat.

About the quantum corrections to the FRW model, a semi-classical approach is proposed with the Hamiltonian formulation \cite{Bojowald2001b}. There are mainly two expected types of quantum corrections from the Hamiltonian: holonomy \cite{Singh2005,Chiou2009,Chiou2009a} and inverse volume \cite{Bojowald2001c}. One can obtain the equations of motion of the connection variables (known as the Ashtekar variables in LQC) by calculating their Poisson brackets with the Hamiltonian. In this framework, the scale factor is governed by the quantum corrected Friedmann equation \cite{Bojowald2005,Vandersloot2005}.
The cosmological inflation is still required to solve the problems such as the horizon problem (pointed out by Charles Misner in the late 1960s), the flatness problem (first proposed by Robert Dicke in 1969), the monopole problem \cite{Zeldovich1978,Preskill1979}, and others \cite{Guth1981}. Fortunately, cosmological inflation occurs naturally after the quantum bounce due to the existence of a scalar field \cite{Bojowald2002}. 

The primordial perturbations in our universe within the framework of LQC are expected to be different from those within the standard cosmology \cite{Bojowald2006,Mielczarek2007,Bojowald2007,Bojowald2007a}. In particular we are interested in the tensor mode, which is known as gravitational waves (GWs). The power spectrum of such primordial GWs in LQC has been calculated in previous studies using analytical and numerical approaches \cite{Mielczarek2008a,Mielczarek2009,Mielczarek2012}. Those studies used the Bogoliubov transformations to evolve the GWs in LQC. The features of such primordial GWs may be observable in the B-mode polarization of the Cosmic Microwave Background (CMB) \cite{Barrau2009,Mielczarek2010b}.

In this study, we propose a new formalism that employs the transfer functions to bring arbitrary GWs from any given initial time to a designated final time. This enables us to evolve the GWs from the parent universe through the possible deflation, the quantum bounce, the inflation, and any epoch of our interest. In particular, we discuss GWs with the holonomy corrections \cite{Grain2010,Mielczarek2010,Mielczarek2009}. Our formalism is so transparent that we are able to resolve, for example, the IR suppression problem. We also show that our new approach using the transfer functions is equivalent to and produces same results as the Bogoliubov transformations, but with great simplicity and thus much lower cost in calculation. This advance should be beneficial to those studying the GWs in LQC. 

In addition to the transfer-function formalism, in this work we shall propose the `field-free approximation' for the effective mass in the quantum bounce epoch, which in turn dramatically improves the accuracy in the predicted GW power spectrum.

Here is the layout of the paper.
Section \ref{sec02} provides and discusses the background dynamics of cosmic evolution within the context of the LQC. Section \ref{sec03} introduces the new formalism that employs transfer functions. 
Section \ref{sec04} and \ref{sec05} calculate the transfer functions for all related epochs of cosmic evolution. In particular the `field-free approximation' is proposed to improve on the handling of the effective mass in quantum bounce.
Section \ref{sec06} verifies the consistency with the Bogoliubov transformations and considers additional but importantly related issues, including resolving the IR suppression problem and improving the accuracy of the predicted GW power spectrum using the field-free approximation.
Finally, Section \ref{sec07} summarizes the work and gives conclusions.
The units of all physical quantities appearing in this paper are normalized to the Planckian units ($c=G=\hbar=k_B\equiv1$) except for those specially labeled.

% -----------------------------------------------------------------------
\section{Cosmological background dynamics}\label{sec02}

We start by considering the effective Hamiltonian of background dynamics in LQC with $n$th-order holonomy corrections in terms of the connection variables $p$ and $c$ \cite{Chiou}:
\begin{align}
H^{(n)}_{\bar{\mu}} & = H^{(n)}_{\rm grav} + H_\phi \nonumber \\
					& = -\frac{3}{8\pi\gamma^2} (c_{\rm h}^{(n)})^2 \sqrt{p} + \frac{\pi_\phi^2}{2p^{3/2}} + p^{3/2} \frac{m_\phi^2 \phi^2 }{2},
\end{align}
where the Chaotic potential $V(\phi)=m_\phi^2\phi^2/2$ has been used with a scalar field $\phi$, its conjugate momentum $\pi_\phi$, and its mass $m_\phi$, which has been chosen as $10^{-6}$ throughout this paper;
the Barbero-Immirzi parameter \cite{BarberoG.1995,Immirzi1997} $\gamma=\log(3)/\sqrt{2}\pi$ can be obtained from the computation of the entropy of black holes \cite{Meissner2004a};
the $n$th-order holonomized connection variable $c_{\rm h}^{(n)}$ is defined by \cite{Chiou2009}
\begin{align}\label{eq02}
c_{\rm h}^{(n)} \equiv \frac{1}{\bar{\mu}}\sum_{k=0}^n \frac{(2k)!}{2^{2k}(k!)^2(2k+1)}(\sin\bar{\mu}c)^{2k+1};
\end{align}
a discreteness variable $\bar{\mu}=\sqrt{\Delta/p}$ is defined in Ref.~\cite{Ashtekar2006b}, with $\Delta=2\sqrt{3}\pi\gamma$ being the standard choice of the area gap in the full theory of LQG.

The connection variables are related to the cosmic scale factor $a$ and the Hubble parameter $H$ by \cite{Bojowald2005}
\begin{align}
|p| = \frac{1}{4}L^2a^2 = \frac{1}{4}a^2, \quad
c = \frac{1}{2}L(K + \gamma aH) = \frac{1}{2}\gamma aH.
\end{align}
They satisfy the canonical relation \cite{Bojowald2005}
\begin{align}
\left[c,p\right]_{\rm PB}=\frac{8\pi\gamma}{3}.
\end{align}
The subscript `PB' denotes that the calculating rule follows `Possion Bracket' rather than the commutation. The curvature parameter $K$ of the FRW metric vanishes as we consider a flat universe here. 
We have chosen the coordinate length of the finite-sized cubic cell $L$ in LQG to be unity.
The scalar field and its conjugate momentum satisfy the canonical relation \cite{Chiou2009}
\begin{align}
\left[\phi, \pi_\phi\right]_{\rm PB}=1.
\end{align}

By imposing the condition of the Hamiltonian constraint \cite{Thiemann1998}
\begin{align}\label{eq07}
H^{(n)}_{\bar{\mu}} = 0,
\end{align}
we may have a lapse function $N$ for changing the time variable from the coordinate (proper) time $t$ to a new parametric time $t'$ via $dt'=N^{-1}dt$. The corresponding Hamiltonian with the new time variable is therefore $H^{(n)'}_{\bar{\mu}}=NH^{(n)}_{\bar{\mu}}$. The new time variable $t'$ in the case $N=\sqrt{p}$ corresponds to the conformal time $\eta$ defined by $d\eta=a^{-1}dt$.
The four equations of motion can then be obtained as
\begin{align}
\frac{dq}{dt} &= [q,H^{(n)}_{\bar{\mu}}]_{\rm PB},
\end{align}
where $q$ represents $p$, $c$, $\phi$, or $\pi_\phi$ \cite{Grain2010}.

In the simplest prescription of the holonomy corrections $n=0$, the connection variable reduces to \cite{Chiou2009,Chiou2009a}
\begin{align}
c_{\rm h}^{(0)} = \frac{\sin\bar{\mu}c}{\bar{\mu}}.
\end{align}
The total Hamiltonian becomes \cite{Mielczarek2010}
\begin{align}
H^{(0)}_{\bar{\mu}} = -\frac{3}{8\pi\gamma^2} \frac{\sin^2\bar{\mu}c}{\bar{\mu}^2} \sqrt{p} + \frac{\pi_\phi^2}{2p^{3/2}} + p^{3/2} \frac{m_\phi^2 \phi^2 }{2}.
\end{align}
Thus the energy density of the scalar field should be \cite{Mielczarek2010}
\begin{align}
\rho_\phi = \frac{\pi_\phi^2}{2p^3} + \frac{m_\phi^2 \phi^2 }{2} = \frac{3}{8\pi\gamma^2\Delta}\sin^2\bar{\mu}c,
\end{align}
in order to satisfy the Hamiltonian constraint. We can see that a quantum bounce occurs when $\rho_\phi$ reaches its maximum, which we call the critical energy density \cite{Chiou2009}
\begin{align}
\rho_c = \frac{\sqrt{3}}{16\pi^2\gamma^3}.
\end{align}
In the meantime,
\begin{align}\label{eq13}
c(t=0) = \frac{\pi}{2}\sqrt{\frac{p(t=0)}{\Delta}}.
\end{align}
We set the proper time $t$ to be zero at the quantum bounce. With Eq.~\eqref{eq07} and Eq.~\eqref{eq13}, and choosing the scale factor to be unity at $t=0$, we already have three conditions for solving the four coupled equations of motion. The last condition we need is the relation between the scalar field and its conjugate momentum at $t=0$.

As considered in the literature, the relation between the scalar field and its conjugate momentum (more precisely, the ratio between the potential and the kinetic energy of the scalar field) at $t=0$ determines the time-symmetry of cosmological background with respect to the quantum bounce \cite{Mielczarek2010,Bolliet2015}. As mentioned above, $\rho_\phi=\rho_c$ is satisfied at $t=0$. We have 
\begin{align}
\frac{1}{2}\dot{\phi}^2 + \frac{1}{2}m_\phi^2\phi^2 = \rho_c. \qquad \left(t=0\right)
\end{align}
Now we define a new parameter $\theta_B$ by
\begin{equation}\label{eq14}
\begin{split}
\cos\theta_B = \frac{1}{\sqrt{2\rho_c}}\dot{\phi} \,, \quad
\sin\theta_B = \frac{1}{\sqrt{2\rho_c}}m_\phi\phi \,,
\quad
\left(t=0\right)
\end{split}
\end{equation}
so that the time-symmetry of cosmological background is simply described by $\theta_B$.

Theoretically the value of $\theta_B$ lies within the interval $[0,\pi/2]$. The special case $\theta_B=0$ corresponds to a time-symmetric cosmological background with $\phi=0$ at $t=0$ with a Chaotic inflation. In the literature asymmetric backgrounds are considered. For example the `shark-fin type' background \cite{Mielczarek2010} gives a relatively large expansion ratio of the universe through the quantum bounce and inflation and thus in turn helps solving the cosmological problems such as the horizon problem and the flatness problem.

In the framework of LQC, the singularity problem of the standard cosmology is avoided by introducing the quantum bounce. This allows for the existence of the pre-bouncing phases and thus the so-called `parent universe'. 
To provide a complete recipe for evolving the GWs in LQC,
in the next sessions we shall consider all the related stages of cosmic evolutionary history: pre-deflationary contraction, deflation, quantum bounce, inflation, and post-inflationary expansion.

% -----------------------------------------------------------------------
\section{New approach to gravitational-wave evolution}\label{sec03}
\subsection{Overview on modern research}

In the standard cosmological model the large-scale cosmic structures are originated from the primordial perturbations created through the quantum fluctuations during inflation. The inflationary process froze out the pair-created gravitons and thus generated the GWs as the by-product of inflation. 
The standard power spectrum of these primordial GWs on sub-horizon scales in a de Sitter expansion is
\begin{align}\label{eq16}
P_T^{\rm std} = \frac{8\pi H^2}{k^3},
\end{align}
where
$k$ is the wavenumber,
and $H$ is supposed to be constant during inflation.

Recent studies considered the primordial GWs generated during the quantum bounce in LQC \cite{Grain2010,Mielczarek2010} but not earlier.
However, there could be GWs even in the parent universe surviving the quantum bounce and manifesting themselves today with different features as compare with those studied before. Thus one of the major contributions of our work is to provide a new framework that enables us to evolve the `pre-existing GWs' from the parent universe through the quantum bounce to the end of inflation and even to the present.

\subsection{Gravitational waves in loop quantum cosmology}

The evolution of GWs in LQC can be described by the wave equation \cite{Bojowald2007a}
\begin{align}\label{eq17}
\frac{d^2}{d\eta^2} h_i^j + 2H \frac{d}{d\eta} h_i^j + (-\bigtriangledown^2 + m_Q^2)h_i^j = 0.
\end{align}
The GW functions are given by $h_i^j=h_i^j(\eta,\textbf{x})$ with $h_1^1 = -h_2^2 = h_\oplus$ and $h_2^1 = h_1^2 = h_\otimes$ in the transverse-traceless gauge. They are decomposed from the metric perturbations. The holonomy correction term $m_Q$ is given by \cite{Bojowald2007a}
\begin{align}\label{eq18}
m_Q^2 = 16\pi a^2 \frac{\rho}{\rho_c}(\frac{2}{3}\rho - V),
\end{align}
which vanishes in a classical regime.

It is clear that there exists a damping term in the wave equation. It means that GWs will be diluted by the cosmic expansion. Therefore, it is more convenient to solve the wave equation and understand the physics by defining a new comoving function $u$ \cite{Mielczarek2010}:
\begin{align}
u = u(\textbf{x},\eta) = \frac{ah_\oplus}{\sqrt{16\pi}} = \frac{ah_\otimes}{\sqrt{16\pi}}.
\end{align}
In the Fourier space,
the wave equation then becomes \cite{Mielczarek2010}:
\begin{equation}\label{eq20}
\frac{d^2}{d\eta^2} \tilde{u}_\textbf{k}(\eta) + (k^2 + m_{\rm eff}^2)\tilde{u}_\textbf{k}(\eta) = 0,
\end{equation}
where
\begin{align}
\tilde{u}_\textbf{k} = \tilde{u}_\textbf{k}(\eta) = \int \frac{d^3k}{(2\pi)^3} u(\textbf{x},\eta) e^{-i\textbf{k} \cdot \textbf{x}},
\end{align}
and the effective mass is defined as
\begin{align}\label{eq21}
m_{\rm eff}^2 = m_Q^2 - \frac{1}{a} \frac{d^2a}{d\eta^2}.
\end{align}

\begin{figure}[t!]
	\centering
	\includegraphics[width=0.48\textwidth]{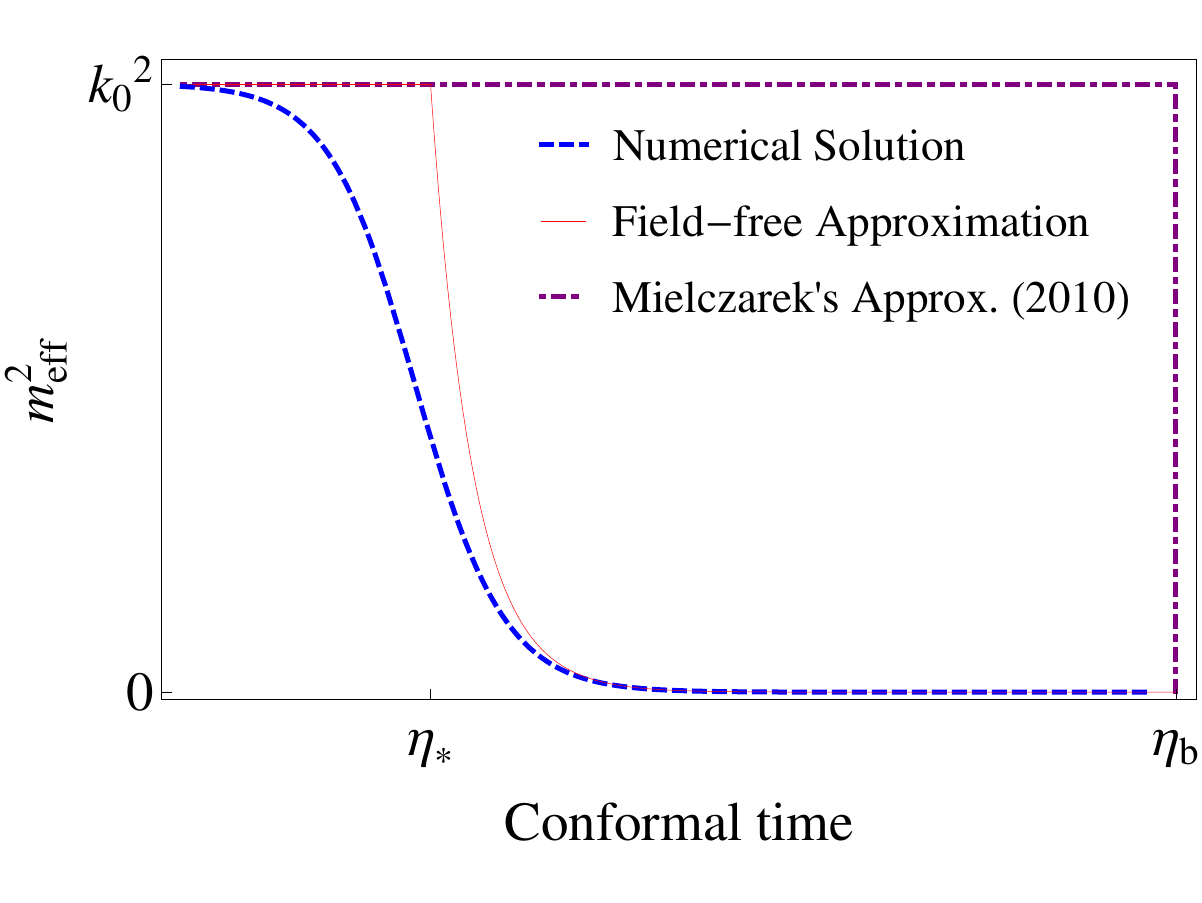}
	\caption{\label{fig01}The effective mass as a function of the conformal time, evolving from the quantum bounce ($\eta=0$) to the beginning of inflation ($\eta=\eta_{\rm b}$).}
\end{figure}

Figure \ref{fig01} shows the effective mass as a function of the conformal time. The part for $\eta<0$ is not shown as it is expected to be almost time-symmetric with respect to $\eta=0$.
The blue-dashed line is our numerical result, indicating the inaccuracy of other approximations which we shall discuss later. The effective mass is relatively large around the quantum bounce, meaning that the quantum effects acting on GWs become significant near the quantum bounce. In addition, Eq.~\eqref{eq20} also tells us that the quantum effects behave much more obviously on larger scales (smaller $k$). This feature provides critical insight into the observational tests for such quantum effects at the bounce.

Such scale-dependent features can also be argued from a different perspective through the comoving coordinates. The wavenumber $k$ is related to the comoving wavelength $\lambda_k$ of the GW as $\lambda_k\sim 2\pi /k$. According to Eq.~\eqref{eq20}, the GWs reduce to plane waves on small scales.
On the other hand, the comoving Hubble radius goes to infinity at the quantum bounce, which is the turning point ($H=0$) of the cosmic contraction. It means that the whole universe is in causal connection due to strong quantum effects at the bounce. For a GW originally on super-horizon scales evolving through the quantum bounce, it will suddenly oscillate after the horizon entry near the bounce. Therefore, we expect that the large-scale power spectrum in such scenarios to be potentially different from that in the standard model.

\subsection{The transfer functions}

Eq.~\eqref{eq20}, which governs the evolution of the GWs, is a linear second-order ordinary differential equation of $\tilde{u}_\textbf{k}(\eta)$, so for any given set of initial and final times there must exist certain transfer functions connecting any arbitrarily given initial conditions with their final solutions.
In other words, if we define
\begin{align}
\tilde{\bf U}_\textbf{k} (\eta) \equiv
\begin{bmatrix}
\tilde{u}_\textbf{k}(\eta') \\[3pt]
\frac{d \tilde{u}_\textbf{k}(\eta')}{d\eta'}
\end{bmatrix}_{\eta'=\eta},
\end{align}
then the above statement goes as
\begin{align}\label{eq22}
\tilde{\bf U}_\textbf{k}(\eta_{\rm f}) = 
{\bf T}_\textbf{k}(\eta_{\rm i},\eta_{\rm f})
\tilde{\bf U}_\textbf{k}(\eta_{\rm i}),
\end{align}
where 
$\tilde{\bf U}_\textbf{k}(\eta_{\rm i})$ is the initial condition at $\eta_{\rm i}$, 
$\tilde{\bf U}_\textbf{k}(\eta_{\rm f})$ is the final state at $\eta_{\rm f}$,
and ${\bf T}_\textbf{k}(\eta_{\rm i},\eta_{\rm f})$ is a $2\times 2$ transfer matrix that contains four independent transfer functions:
\begin{align}\label{eq23}
{\bf T}_\textbf{k}(\eta_{\rm i},\eta_{\rm f}) =
\left[
\begin{array}{cc}
T(\textbf{k})_{11} & T(\textbf{k})_{12}\\
T(\textbf{k})_{21} & T(\textbf{k})_{22}\\
\end{array}
\right].
\end{align}
These four transfer functions are 
independent of the initial condition and are
purely determined by the evolution equation, whose form changes in different epochs following different cosmological background dynamics.

If the evolutionary history of the universe can be divided into $N$ epochs in sequence, namely $[\eta_{\rm i}^x,\eta_{\rm f}^x]$ $(x=1...N)$, then the overall transfer matrix ${\bf T}_\textbf{k}(\eta_{\rm i},\eta_{\rm f})$ can be obtained by combining the transfer matrices ${\bf T}^x_\textbf{k}(\eta_{\rm i}^x,\eta_{\rm f}^x)$ of each individual epoch: 
\begin{align}\label{eq24}
	{\bf T}_\textbf{k}(\eta_{\rm i},\eta_{\rm f}) =
	\prod\limits_{x=N}^{1} {\bf T}^x_\textbf{k}(\eta_{\rm i}^x,\eta_{\rm f}^x).
\end{align}
For a GW equation that can be solved analytically, its corresponding transfer functions can be easily derived. In Appendix \ref{app03}, we use a toy example to demonstrate how we derive the transfer functions from the analytical solutions of a linear equation of motion.

Now we demonstrate how to numerically obtain the transfer matrix of given $\eta_{\rm i}$ and $\eta_{\rm f}$.
In principle we could numerically evolve any arbitrary initial condition $\tilde{\bf U}_\textbf{k}(\eta_{\rm i})$ to obtain its final state $\tilde{\bf U}_\textbf{k}(\eta_{\rm f})$, and then derive the corresponding transfer functions, which should be valid for any initial condition.
As shown in Eq.~\eqref{eq23}, ${\bf T}_\textbf{k}(\eta_{\rm i},\eta_{\rm f})$ is a $2\times 2$ matrix with four transfer functions, so we need four equations to derive them.
Eq.~\eqref{eq22} actually provides two `independent' equations, so what we need to do is simply to evolve two different initial conditions  
$\tilde{\bf U}_\textbf{k}^{{\rm i} (a)} \equiv 
\tilde{\bf U}_\textbf{k}^{(a)}(\eta_{\rm i})$ ($a=1$, $2$) into their final states $\tilde{\bf U}_\textbf{k}^{{\rm f} (a)} \equiv \tilde{\bf U}_\textbf{k}^{(a)}(\eta_{\rm f})$, 
and then we are ready to solve for the four transfer functions using:
\begin{subequations}
\begin{align}
\tilde{u}_\textbf{k}^{{\rm f}(a)}= T(\textbf{k})_{11}\tilde{u}_\textbf{k}^{{\rm i}(a)} + T(\textbf{k})_{12}\frac{d\tilde{u}_\textbf{k}}{d\eta}^{{\rm i}(a)}, \\
\frac{d\tilde{u}_\textbf{k}}{d\eta}^{{\rm f}(a)} = T(\textbf{k})_{21}\tilde{u}_\textbf{k}^{{\rm i}(a)} + T(\textbf{k})_{22}\frac{d\tilde{u}_\textbf{k}}{d\eta}^{{\rm i}(a)},
\end{align}
\end{subequations}
where $a=1$, $2$.
The explicit forms of the solved transfer functions are:
\begin{subequations}\label{Tkii}
\begin{align}
T(\textbf{k})_{11} &= \frac{\tilde{u}_\textbf{k}^{\rm f(1)}\frac{d\tilde{u}_\textbf{k}}{d\eta}^{\rm i(2)}-\tilde{u}_\textbf{k}^{\rm f(2)}\frac{d\tilde{u}_\textbf{k}}{d\eta}^{\rm i(1)}}{\tilde{u}_\textbf{k}^{\rm i(1)}\frac{d\tilde{u}_\textbf{k}}{d\eta}^{\rm i(2)}-\tilde{u}_\textbf{k}^{\rm i(2)}\frac{d\tilde{u}_\textbf{k}}{d\eta}^{\rm i(1)}};
\\[3pt]
T(\textbf{k})_{12} &= \frac{\tilde{u}_\textbf{k}^{\rm f(1)}\tilde{u}_\textbf{k}^{\rm i(2)}-\tilde{u}_\textbf{k}^{\rm f(2)}\tilde{u}_\textbf{k}^{\rm i(1)}}{\frac{d\tilde{u}_\textbf{k}}{d\eta}^{\rm i(1)}\tilde{u}_\textbf{k}^{\rm i(2)}-\frac{d\tilde{u}_\textbf{k}}{d\eta}^{\rm i(2)}\tilde{u}_\textbf{k}^{\rm i(1)}};
\\[1pt]
T(\textbf{k})_{21} &= \frac{\frac{d\tilde{u}_\textbf{k}}{d\eta}^{\rm f(1)}\frac{d\tilde{u}_\textbf{k}}{d\eta}^{\rm i(2)}-\frac{d\tilde{u}_\textbf{k}}{d\eta}^{\rm f(2)}\frac{d\tilde{u}_\textbf{k}}{d\eta}^{\rm i(1)}}{\tilde{u}_\textbf{k}^{\rm i(1)}\frac{d\tilde{u}_\textbf{k}}{d\eta}^{\rm i(2)}-\tilde{u}_\textbf{k}^{\rm i(2)}\frac{d\tilde{u}_\textbf{k}}{d\eta}^{\rm i(1)}};
\\[1pt]
T(\textbf{k})_{22} &= \frac{\frac{d\tilde{u}_\textbf{k}}{d\eta}^{\rm f(1)}\tilde{u}_\textbf{k}^{\rm i(2)}-\frac{d\tilde{u}_\textbf{k}}{d\eta}^{\rm f(2)}\tilde{u}_\textbf{k}^{\rm i(1)}}{\frac{d\tilde{u}_\textbf{k}}{d\eta}^{\rm i(1)}\tilde{u}_\textbf{k}^{\rm i(2)}-\frac{d\tilde{u}_\textbf{k}}{d\eta}^{\rm i(2)}\tilde{u}_\textbf{k}^{\rm i(1)}}.
\end{align}
\end{subequations}
This provides a complete recipe for obtaining accurate numerical results.
We always cross check our numerical results with the analytical results, wherever the latter are available. The numerical results are also useful for verifying our approximations where exact analytical solutions are not available. 
Because of the cosmological principle, we shall drop the directional dependence of $\textbf{k}$ to consider only $k\equiv |\textbf{k}|$ in the following discussions.

% -----------------------------------------------------------------------
\section{Transfer functions for inflationary and deflationary epochs}\label{sec04}
\subsection{The slow-roll inflation}\label{sec4a}

The definition of inflation is simply that the universe undergoes an era of accelerated expansion (see Ref.~\cite{AndrewR.Liddle2000} for a good review):
\begin{align}\label{eq32}
\ddot{a}>0.
\end{align}
After the quantum bounce there are two periods when inflation occurs: the cosmological inflation in the standard cosmology and the super-inflationary phase near the quantum bounce. In the following, we focus on the former,  which is necessary for solving the cosmological conundrums.

To investigate the inflationary epoch, we adopt the `potential slow-roll approximation' (PSRA) \cite{Steinhardt1984,Liddle1992}. 
Under the PSRA, the Chaotic inflation may be approximated as the `de Sitter expansion', when the scale factor grows exponentially so that the Hubble parameter stays as a constant $H_{\rm dS}$.
The scale factors at the beginning ($\eta =\eta_{\rm b}$) and the end ($\eta =\eta_{\rm e}$) of inflation are denoted as $a_{\rm b}\equiv a(\eta_{\rm b})$ and $a_{\rm e}\equiv a(\eta_{\rm e})$ respectively.
The definition of conformal time (see e.g.~Ref.~\cite{Dodelson2003}) thus gives
\begin{align}\label{eq37}
\eta' = \int_{a_{\rm e}}^a \frac{da}{Ha^2} = \frac{1}{H_{\rm dS}}\int_{a_{\rm e}}^a \frac{da}{a^2} = \frac{-1}{aH_{\rm dS}} + \frac{1}{a_{\rm e}H_{\rm dS}}.
\end{align}
We note that $\eta'$ is originated at the end of inflation while $\eta$ is originated at the quantum bounce, so $\eta'$ is negative during inflation. These are related as: 
\begin{align}\label{eq37.1}
\eta'=\eta-\eta_{\rm e}.
\end{align}
For more review on inflationary cosmology please refer to Ref.~\cite{Jiun-HueiProtyWu1996} and Ref.~\cite{AndrewR.Liddle2000}.

\subsection{Analytical solutions with de Sitter expansion}\label{sec4b}

In this epoch the quantum correction term $m_{\rm Q}^2$ in Eq.~\eqref{eq17} vanishes, therefore Eq.~\eqref{eq20} becomes
\begin{align}\label{eq37.5}
\frac{d^2}{d\eta^2} \tilde{u}_k + \left(k^2 - \frac{1}{a}\frac{d^2a}{d\eta^2}\right) \tilde{u}_k = 0.
\end{align}
In the regime of de Sitter expansion, Eq.~\eqref{eq37.5} can be rewritten as:
\begin{equation}\label{eq38}
\begin{split}
\frac{d^2}{d\eta_{\rm I}^2} \tilde{u}_k + \left(k^2-\frac{2}{\eta_{\rm I}^2}\right) \tilde{u}_k = 0 \,,
\quad
\eta_{\rm I}\equiv \frac{1}{a_{\rm e}H_{\rm dS}} - \eta', % modified HERE
\end{split}
\end{equation}
where `$\eta_{\rm I}$' denotes the conformal time defined in the inflationary epoch.
It is obvious that Eq.~\eqref{eq38} is in the form of a Bessel differential equation of order $3/2$, so its solution should be simply the linear sum of Bessel functions:
\begin{align}
\tilde{u}^{\rm I}_k(\eta_{\rm I}) = &A^{\rm I} \sqrt{\eta_{\rm I}}J_{3/2}(k\eta_{\rm I}) + B^{\rm I}\sqrt{\eta_{\rm I}}J_{-3/2}(k\eta_{\rm I}),
\end{align}
where $J$ is the Bessel function of the first kind.
Here we use the superscript `I' to denote quantities specifically for the inflationary epoch.
The coefficients $A^{\rm I}$ and $B^{\rm I}$ may be determined by the initial conditions. More details about the mathematics of Bessel differential equation of order $3/2$ are provided in Section \ref{app02}, and the transfer functions corresponding to Eq.~\eqref{eq38} are presented as Eq.~\eqref{eqb12.0} in Section \ref{app02}.

\begin{figure*}[t!]
	\centering
	\includegraphics[width=0.96\textwidth]{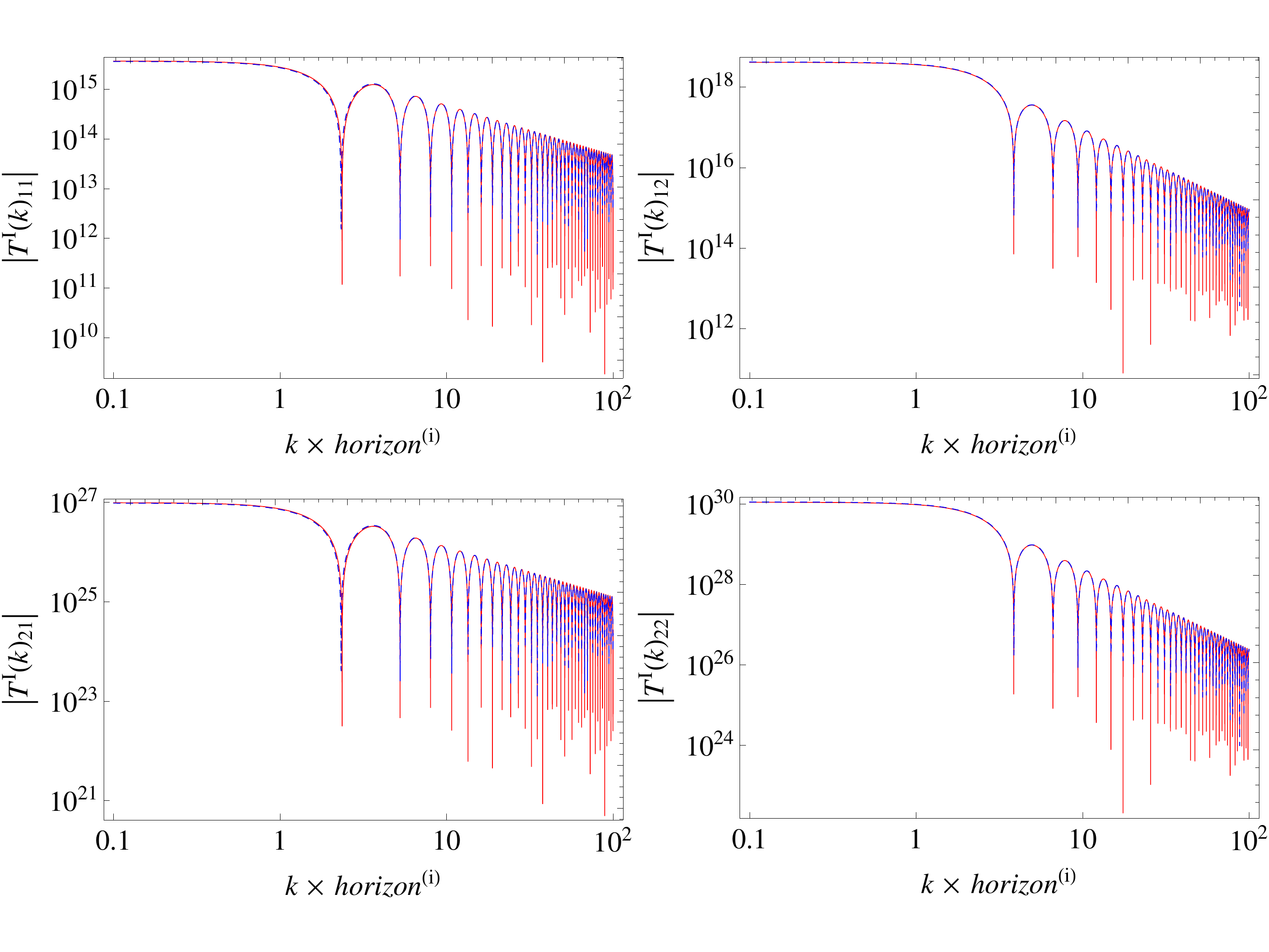}
	\caption{\label{fig02}
		The numerical (blue-dashed) and analytical (red-solid) results for the transfer functions of GWs in de Sitter inflation with Chaotic potential. 
		The consistency between the two is obvious.}
\end{figure*}

By taking the limit $a_{\rm e}\gg a_{\rm b}$, the four transfer functions as defined in Eqs.~\eqref{eq22} and \eqref{eq23} for the inflationary epoch can be obtained as:
\begin{subequations}\label{eq44}
\begin{align}\label{eq44.1}
T^{\rm I}(k)_{11} &= \frac{a_{\rm e}H_{\rm dS}}{k}[\sin(k\Delta\eta_{\rm I})-j_1(k\Delta\eta_{\rm I})],
\\[2pt]
T^{\rm I}(k)_{12} &= \frac{a_{\rm e}H_{\rm dS}}{k^2}[\cos(k\Delta\eta_{\rm I})-j_0(k\Delta\eta_{\rm I})],
\\[1pt]
T^{\rm I}(k)_{21} &= -\frac{a_{\rm e}^2H_{\rm dS}^2}{k}[\sin(k\Delta\eta_{\rm I})-j_1(k\Delta\eta_{\rm I})],
\\[1pt]
T^{\rm I}(k)_{22} &= -\frac{a_{\rm e}^2H_{\rm dS}^2}{k^2}[\cos(k\Delta\eta_{\rm I})-j_0(k\Delta\eta_{\rm I})],
\end{align}
\end{subequations}
where $\Delta\eta_{\rm I}\equiv\eta_{\rm e}-\eta_{\rm b}$ is the duration of inflation, and the $j$'s are the spherical Bessel functions of the first kind. These four transfer functions are shown in Figure \ref{fig02}, where such analytical results are numerically verified using Eq.\eqref{Tkii} at high accuracy.

Although these transfer functions are obtained in a purely classical manner, they are consistently useful in a quantum mechanical problem. Considering the quantum fluctuations in the Minkowski spacetime (see e.g. Ref.~\cite{Dodelson2003})
\begin{equation}
\begin{split}
\tilde{u}_k(\eta_{\rm b})=\frac{1}{\sqrt{2k}}e^{-ik\eta_{\rm b}} \,
\quad \mbox{and} \quad
\frac{d\tilde{u}_k}{d\eta}(\eta_{\rm b})=\frac{-ik}{\sqrt{2k}}e^{-ik\eta_{\rm b}},
\end{split}
\end{equation}
with the sub-horizon (large-$k$) limit
\begin{align}\label{eq43}
\lim_{k\rightarrow\infty} {\bf T}^{\rm I}&(k) = 
\begin{bmatrix}
  \frac{a_{\rm e}H_{\rm dS}}{k}\sin(k\Delta\eta_{\rm I}) & \frac{a_{\rm e}H_{\rm dS}}{k^2}\cos(k\Delta\eta_{\rm I})
  \\[3pt]
  -\frac{a_{\rm e}^2H_{\rm dS}^2}{k}\sin(k\Delta\eta_{\rm I}) & -\frac{a_{\rm e}^2H_{\rm dS}^2}{k^2}\cos(k\Delta\eta_{\rm I})
\end{bmatrix},
\end{align}
we can obtain the primordial tensor power spectrum at the end of inflation as
\begin{align}\label{PT-sub}
\lim_{k\rightarrow \infty} P_{\rm T}^{\rm dS}(k) = \frac{16\pi}{a_{\rm e}^2}|\tilde{u}_k(\eta_{\rm e})|^2 = \frac{8\pi H_{\rm dS}^2}{k^3} = P_{\rm T}^{\rm std}(k).
\end{align}
This shows the consistency with the result from quantizing the GWs considered before.

In summary, we conclude that the transfer functions successfully classicalize a quantum mechanical problem. The quantization process and the Bogoliubov transformations are now substituted by simple algebraic calculations. 
In the later section we shall provide further consistency check to show the power of the transfer functions.

\subsection{A possible deflationary epoch in the parent universe}\label{app01}

If there exists a parent universe, there must be a contractive process before reaching the limit to trigger the quantum bounce.
We thus consider in this process a possible `deflation', which shares the same mathematical definition with inflation as in Eq.~\eqref{eq32}.
In addition to the primordial perturbations, which are originated from the quantum bounce and inflation, in this paper we also derive the transfer functions for the deflationary epoch, extending the framework in literature well through the bounce backwards in time into the parent universe.

To handle deflation, we also employ the PSRA. Again the Hubble parameter is expected to vary so slowly and thus well approximated by the `de Sitter contraction', where the scale factor decays exponentially so that the Hubble parameter stays as a negative constant $H_{\rm dS}^{\rm D}<0$. 
We can then define the conformal time with an origin at the beginning of deflation as
\begin{align}\label{eqa1}
\eta_{\rm D}' = \int_{a_{\rm b}^{\rm D}}^a \frac{da}{Ha^2} = \frac{1}{H_{\rm dS}^{\rm D}}\int_{a_{\rm b}^{\rm D}}^a \frac{da}{a^2}
= \frac{1}{H_{\rm dS}^{\rm D}} \left(\frac{-1}{a} + \frac{1}{a_{\rm b}^{\rm D}}\right).
\end{align}
Note that $a<a_{\rm b}^{\rm D}$ during deflation so $\eta_{\rm D}'$ is positive, and that
\begin{align}\label{eqa1.1}
	\eta_{\rm D}'=\eta-\eta_{\rm b}^{\rm D}.
\end{align}
The superscript or subscript `D' denotes for deflation. The equation of $\tilde{u}_k$ in the deflationary epoch can now be written as
\begin{equation}\label{eqa2}
\begin{split}
\frac{d^2}{d\eta_{\rm D}^2} \tilde{u}_k + \left(k^2 - \frac{2}{\eta_{\rm D}^2}\right) \tilde{u}_k = 0 \,,
\quad
\eta_{\rm D}\equiv \eta_{\rm D}' -\frac{1}{a_{\rm b}^{\rm D}H_{\rm dS}^{\rm D}} \,.
\end{split}
\end{equation}
This is in the form of a Bessel differential equation of order $3/2$ so its general solution is
\begin{align}\label{eqa3}
\tilde{u}^{\rm D}_k(\eta_{\rm D}) = &A^{\rm D} \sqrt{\eta_{\rm D}} J_{3/2}(k\eta_{\rm D}) + B^{\rm D} \sqrt{\eta_{\rm D}} J_{-3/2}(k\eta_{\rm D}),
\end{align}
where the coefficients $A^{\rm D}$ and $B^{\rm D}$ may be determined by the initial conditions.
The transfer functions ${\bf T}^{\rm D}(k) \equiv {\bf T}^{\rm D}_k(\eta^{\rm D}_{\rm b},\eta^{\rm D}_{\rm e})$ corresponding to Eq.~\eqref{eqa2} are presented as Eq.~\eqref{eqb14.0} in Section \ref{app02}.

As we shall argue in Section \ref{app02}, ${\bf T}^{\rm D}(k)$ should be the inverse matrix of ${\bf T}^{\rm I}(k)$ if the cosmological background is time-symmetric with respect to the quantum bounce. More details about the transfer functions for Bessel differential equation of order 3/2 are provided below.

\subsection{The transfer functions and their symmetric property in de Sitter space}\label{app02}

The Bessel differential equation is a second-order linear ordinary differential equation \cite{Weisstein}:
\begin{align}
x^2\frac{d^2u}{dx^2}+x\frac{du}{dx}+(x^2-n^2)u=0.
\end{align}
A modified expression of the Bessel differential equation of order $n$ is  \cite{Bowman1958,Weisstein}:
\begin{align}\label{eqb1}
\frac{d^2u}{dx^2} - \frac{2\alpha-1}{x}\frac{du}{dx} + \left(\beta^2\gamma^2x^{2\gamma-2} + \frac{\alpha^2-n^2\gamma^2}{x^2}\right)u = 0.
\end{align}
Here the parameters $\alpha$, $\beta$, $\gamma$ and $n$ are purely mathematical, not related to the cosmological context.
The general solutions are therefore \cite{Weisstein}:
\begin{align}\label{eqb2}
u(x) = \left\{
\begin{array}{l}
	x^\alpha\left[AJ_n(\beta x)+BY_n(\beta x)\right] \\[3pt]
	x^\alpha\left[AJ_n(\beta x)+BJ_{-n}(\beta x)\right]
\end{array} \
\begin{array}{l}
	\mbox{for integer} \ n, \\[3pt]
	\mbox{for non-integer} \ n.
\end{array}\right.
\end{align}
By comparing Eq.~\eqref{eq38} and Eq.~\eqref{eqa2} with Eq.~\eqref{eqb1},
we see that both Eq.~\eqref{eq38} and Eq.~\eqref{eqa2} are Bessel differential equations of order $3/2$, and that
\begin{align}\label{eqb3}
\{\alpha,\beta,\gamma,n\} = \left\{\frac{1}{2},k,1,\frac{3}{2}\right\}.
\end{align}
Substituting these numbers into Eq.~\eqref{eqb2} gives the general solutions as
\begin{align}\label{eqb4}
u_k(x) = \sqrt{x}\left[AJ_{3/2}(kx)+BJ_{-3/2}(kx)\right].
\end{align}
The resulting transfer functions ${\bf T}(k) \equiv {\bf T}_k(x_1,x_2)$ that connect the initial state $u_k(x_1)$ with the final state $u_k(x_2)$ are
\begin{widetext}
\begin{subequations}\label{eqb6}
\begin{align}
T(k)_{11} &= F_4(kx_1)F_2(kx_2) - F_3(kx_1)F_1(kx_2),
\\[1pt]
T(k)_{12} &= \frac{1}{k}\left\{F_1(kx_1)F_2(kx_2) - F_2(kx_1)F_1(kx_2)\right\},
\\[3pt]
T(k)_{21} &= k\left\{F_4(kx_1)F_3(kx_2) + F_3(kx_1)F_4(kx_2)\right\},
\\[7pt]
T(k)_{22} &= F_1(kx_1)F_3(kx_2) + F_2(kx_1)F_4(kx_2),
\end{align}
\end{subequations}
\end{widetext}
with
\begin{align}\label{eqb9.1}
\begin{matrix}
F_1(x) \equiv \cos(x)-j_0(x), \\[5pt]
F_2(x) \equiv \sin(x)-y_0(x), \\[5pt]
F_3(x) \equiv \cos(x)+y_1(x), \\[5pt]
F_4(x) \equiv \sin(x)-j_1(x),
\end{matrix}
\end{align}
where the $y$'s are the spherical Bessel function of the second kind.
These results for the transfer functions are exact without using any approximation.

For the de Sitter expansion, where Eq.~\eqref{eq37} is satisfied, we have
\begin{equation}\label{eqb10}
\begin{split}
x_2 = \frac{1}{a_{\rm e}H_{\rm dS}} \,
\quad \mbox{and} \quad
x_1 = x_2 + \Delta\eta_{\rm I} \equiv \chi_1^{\rm I} \,. \\[4pt]
\end{split}
\end{equation}
The transfer functions for the inflationary epoch are therefore
\begin{widetext}
\begin{subequations}\label{eqb12.0}
\begin{align}\label{eqb12.1}
T^{\rm I}(k)_{11} &= F_4(k\chi_1^{\rm I})F_2\left(\frac{k}{a_{\rm e}H_{\rm dS}}\right) - F_3(k\chi_1^{\rm I})F_1\left(\frac{k}{a_{\rm e}H_{\rm dS}}\right),
\\[1pt]
T^{\rm I}(k)_{12} &= \frac{1}{k}\left[F_1(k\chi_1^{\rm I})F_2\left(\frac{k}{a_{\rm e}H_{\rm dS}}\right) - F_2(k\chi_1^{\rm I})F_1\left(\frac{k}{a_{\rm e}H_{\rm dS}}\right)\right],
\\[2pt]
T^{\rm I}(k)_{21} &= k\left[F_4(k\chi_1^{\rm I})F_3\left(\frac{k}{a_{\rm e}H_{\rm dS}}\right) + F_3(k\chi_1^{\rm I})F_4\left(\frac{k}{a_{\rm e}H_{\rm dS}}\right)\right],
\\[2pt]
T^{\rm I}(k)_{22} &= F_1(k\chi_1^{\rm I})F_3\left(\frac{k}{a_{\rm e}H_{\rm dS}}\right) + F_2(k\chi_1^{\rm I})F_4\left(\frac{k}{a_{\rm e}H_{\rm dS}}\right).
\end{align}
\end{subequations}
\end{widetext}
When taking the limit $a_{\rm e}\gg a_{\rm b}$, Eqs.~\eqref{eqb12.0} reduce exactly to Eqs.~\eqref{eq44}.

An obvious fact is that the determinants of ${\bf T}(k)$ and ${\bf T}^{\rm I}(k)$ are both unity
\begin{align}\label{eq9.5}
\det ({\bf T}(k)) = \det ({\bf T}^{\rm I}(k)) = 1.
\end{align}
However, the determinant of the $T^{\rm I}(k)$ in Eqs.~\eqref{eq44} is zero, not unity. This indicates that the linear independence among the transfer functions is broken when taking the limit $a_{\rm e}\gg a_{\rm b}$. Therefore, we should be very careful whenever we employ Eqs.~\eqref{eq44}, which are not generally valid.

For the de Sitter contraction, where Eq.~\eqref{eqa1} is satisfied, we have
\begin{equation}\label{eqb12}
\begin{split}
x_1 = \frac{-1}{a_{\rm b}^{\rm D}H_{\rm dS}^{\rm D}} \,
\quad \mbox{and} \quad
x_2 = x_1 + \Delta\eta_{\rm D} \equiv \chi_2^{\rm D} \,.
\end{split}
\end{equation}
The transfer functions for the deflationary epoch are therefore
\begin{widetext}
\begin{subequations}\label{eqb14.0}
\begin{align}\label{eqb14.1}
T^{\rm D}(k)_{11} &= F_4\left(\frac{-k}{a_{\rm b}^{\rm D}H_{\rm dS}^{\rm D}}\right)F_2(k\chi_2^{\rm D}) - F_3\left(\frac{-k}{a_{\rm b}^{\rm D}H_{\rm dS}^{\rm D}}\right)F_1(k\chi_2^{\rm D}),
\\[1pt]
T^{\rm D}(k)_{12} &= \frac{1}{k}\left[F_1\left(\frac{-k}{a_{\rm b}^{\rm D}H_{\rm dS}^{\rm D}}\right)F_2(k\chi_2^{\rm D}) - F_2\left(\frac{-k}{a_{\rm b}^{\rm D}H_{\rm dS}^{\rm D}}\right)F_1(k\chi_2^{\rm D})\right],
\\[2pt]
T^{\rm D}(k)_{21} &= k\left[F_4\left(\frac{-k}{a_{\rm b}^{\rm D}H_{\rm dS}^{\rm D}}\right)F_3(k\chi_2^{\rm D}) + F_3\left(\frac{-k}{a_{\rm b}^{\rm D}H_{\rm dS}^{\rm D}}\right)F_4(k\chi_2^{\rm D})\right],
\\[2pt]
T^{\rm D}(k)_{22} &= F_1\left(\frac{-k}{a_{\rm b}^{\rm D}H_{\rm dS}^{\rm D}}\right)F_3(k\chi_2^{\rm D}) + F_2\left(\frac{-k}{a_{\rm b}^{\rm D}H_{\rm dS}^{\rm D}}\right)F_4(k\chi_2^{\rm D}).
\end{align}
\end{subequations}
\end{widetext}

In a cosmological background that is time-symmetric with respect to the quantum bounce,
it is natural to expect that the deflation and inflation are also time-symmetric:
\begin{equation}\label{eqb15}
\begin{split}
\Delta\eta_{\rm I} = \Delta\eta_{\rm D} \,
\quad \mbox{and} \quad
a_{\rm e}H_{\rm dS} = -a_{\rm b}^{\rm D}H_{\rm dS}^{\rm D} \,.
\end{split}
\end{equation}
In such cases, we expect the transfer matrices for inflation and deflation to satisfy
\begin{align}\label{eqb17}
{\bf T}^{\rm I}(k) {\bf T}^{\rm D}(k) = {\bf T}^{\rm D}(k)  {\bf T}^{\rm I}(k) = {\bf I},
\end{align}
where ${\bf I}$ is a $2\times 2$ identity matrix. 
This is indeed verified by our results of transfer matrices.
Eq.~\eqref{eqb17} in turn implies that the evolution of GWs in these two stages behaves exactly as time-reversed with each other.

% -----------------------------------------------------------------------
\section{Transfer functions for quantum bounce epoch}\label{sec05}
\subsection{Definition of the quantum bounce epoch}

The Minkowski vacuum comes before the quantum bounce epoch, which is the first stage for the primordial perturbations to travel through. 
In this work we define the quantum bounce epoch to begin when the quantum effect of the spacetime becomes considerable (the quantum gravity regime) and then to end at the beginning of inflation.

In Eq.~\eqref{eq02}, the discreteness variable $\bar{\mu}$ is directly related to the holonomy corrections, so we could use $\bar{\mu}c_{\rm h}$ to quantify the quantum gravity effect. Thus the quantum gravity regime can be defined as the period when
\begin{align}
0.01 < \bar{\mu}c_{\rm h} < \pi-0.01.
\end{align}
The conformal time at the beginning of the quantum gravity regime $\eta_{\rm i}$ can then be defined through
\begin{align}
\bar{\mu}(\eta_{\rm i})c_{\rm h}(\eta_{\rm i}) = \pi-0.01.
\end{align}
We note that $\bar{\mu}c_{\rm h}$ goes to zero when $t\gg 0$ with the Ashtekar variables reduced to the cosmological parameters in the standard cosmology. Although the above definition for the quantum gravity regime is lack of strict physical justification, it is well motivated enough and convenient for us to discuss the transfer functions in this paper.

In our definition, 
the quantum bounce epoch may contain a classical regime between the quantum gravity regime and the inflationary epoch. 
In summary, the evolutionary history before the end of inflation is now divided into the following stages (in temporal order): the Minkowski vacuum, the quantum bounce epoch, and the inflationary epoch.

\subsection{The field-free approximation for the effective mass}\label{sec5b}

In our framework, a decisive factor for the evolution of GWs is the behavior of the effective mass in Eq.~\eqref{eq20}. During the quantum bounce epoch, the quantum correction term $m_{\rm Q}^2$ is critical and the effective mass depends on both the background dynamics and the scalar potential, as indicated in Eqs.~\eqref{eq18} and \eqref{eq21}.
In a previous study by Mielczarek \textit{et al.}~\cite{Mielczarek2010}, the effective mass is conveniently assumed to be a constant ($k_0$) during the entire quantum bounce epoch, as shown by the purple dot-dashed line in Figure \ref{fig01}. However, it is largely discrepant from the numerical result, as presented by the blue dashed curve. Thus we propose a new approximation for the effective mass in this section, and will compare its results with literature in Section \ref{sec6a}.

We approximate the effective mass in the following way. The simplest case is to consider a scalar-field free cosmological background during the quantum bounce epoch. An exact form of such an effective mass was derived as a function of proper time in Ref.~\cite{Mielczarek2009}. Its corresponding `approximated' function of the conformal time was then provided but only in the classical limit where $|t|\rightarrow\infty$ \cite{Mielczarek2009}.
Based on this result, we build a new analytical form to approximate the numerical solution of the effective mass:
\begin{align}\label{field-free}
m_{\rm eff}^2(\eta)=\left\{
\begin{array}{lcl}
	k_0^2 ~& \mbox{for} ~& {-\eta_\ast < \eta < \eta_\ast}, \\[3pt]
	\frac{1}{4\eta^2}  ~& \mbox{for} ~& {\eta_{\rm i} \le \eta \le -\eta_\ast \mbox{ and } \eta_\ast \le \eta \le \eta_{\rm b}},
\end{array} \right.
\end{align}
where \cite{Mielczarek2009}
\begin{align}
k_0^2 \equiv m_{\rm eff}^2(0) = \frac{16\pi^2\gamma^2 \Delta \pi_\phi^4}{9(\frac{4}{3}\pi\gamma^2 \Delta \pi_\phi^2)^{5/3}},
\end{align}
and
\begin{align}
\frac{1}{4\eta_\ast^2} = k_0^2,
\end{align}
with $\eta_\ast>0$. 
We call Eq.~\eqref{field-free} as the `field-free approximation', 
which is shown as the red solid curve in Figure \ref{fig01}.

As mathematically argued before, the quantum effect for GWs is negligible on small scales (large $k$) but critical on large scales (small $k$).
Its physical reason is that when a large-scale GW propagates through the quantum bounce, it experiences a longer $m_{\rm eff}^2$-dominated period than a small-scale GW. 
Therefore, the inaccuracy of any approximation for the $m_{\rm eff}^2$ is expected to be larger on large scales (small $k$). 

\subsection{Analytical solutions with the field-free approximation}

With the approximation of Eq.~\eqref{field-free}, the quantum bounce epoch can be further divided into three stages: left, middle, and right, along the axis of conformal time. We shall use the superscripts `L', `M', and `R' respectively to label them.

Under the field-free approximation, the effective mass in the middle stage is a constant. The GWs thus propagate like plane waves:
\begin{align}
\frac{d^2}{d\eta^2} \tilde{u}_k + k_\ast^2 \tilde{u}_k = 0,
\end{align}
where $k_\ast\equiv\sqrt{k^2+k_0^2}$ is the effective wavenumber of the plane waves (simple harmonic oscillation). The general solutions can be obtained as (see Appendix \ref{app03} for details):
\begin{align}
\tilde{u}^{\rm M}_k(\eta) = A^{\rm M}\cos(k_\ast\eta) + B^{\rm M}\sin(k_\ast\eta),
\end{align}
where the coefficients $A^{\rm M}$ and $B^{\rm M}$ may be determined by the initial conditions. Thus the corresponding transfer functions are
\begin{align}\label{eq52}
{\bf T}^{\rm M}(k) &\equiv {\bf T}^{\rm M}_k(-\eta_\ast,\eta_\ast) \nonumber \\[3pt] &=
\begin{bmatrix}
  \cos(2k_\ast\eta_\ast) & \frac{1}{k_\ast} \sin(2k_\ast\eta_\ast) \\
  -k_\ast \sin (2k_\ast\eta_\ast) & \cos(2k_\ast\eta_\ast)
\end{bmatrix}.
\end{align}

In the left and right stages, the equation of $\tilde{u}_k$ is the same because the effective mass is time-symmetric with respect to the quantum bounce. According to Eq.~\eqref{field-free}, we have
\begin{align}
\frac{d^2}{d\eta^2}\tilde{u}_k + \left(k^2+\frac{1}{4\eta^2}\right)\tilde{u}_k = 0.
\end{align}
This is a Bessel equation of order zero.

First in the right stage ($\eta>0$), the general solutions are
\begin{align}\label{eq58}
\tilde{u}^{\rm R}_k(\eta) = A^{\rm R} \sqrt{\eta} J_0(k\eta) + B^{\rm R} \sqrt{\eta} Y_0(k\eta),
\end{align}
where $J$ and $Y$ are
the Bessel functions of the first kind and the second kind respectively,
and coefficients $A^{\rm R}$ and $B^{\rm R}$ may be determined by the initial conditions. The corresponding transfer functions ${\bf T}^{\rm R}(k) \equiv {\bf T}^{\rm R}_k(\eta_\ast,\eta_{\rm b})$ are
\begin{widetext}
\begin{subequations}\label{TR}
\begin{align}
T^{\rm R}(k)_{11} &= \frac{1}{D(\eta_\ast)}\sqrt{\frac{\eta_{\rm b}}{\eta_\ast}}[C_Y(\eta_\ast)J_0(k\eta_{\rm b})-C_J(\eta_\ast)Y_0(k\eta_{\rm b})],
\\[1pt]
T^{\rm R}(k)_{12} &= \frac{\sqrt{\eta_\ast\eta_{\rm b}}}{D(\eta_\ast)}[J_0(k\eta_\ast)Y_0(k\eta_{\rm b})-Y_0(k\eta_\ast)J_0(k\eta_{\rm b})],
\\[3pt]
T^{\rm R}(k)_{21} &= \frac{1}{D(\eta_\ast)}\frac{1}{\sqrt{\eta_\ast\eta_{\rm b}}}[C_Y(\eta_\ast)C_J(\eta_{\rm b})-C_J(\eta_\ast)C_Y(\eta_{\rm b})],
\\[1pt]
T^{\rm R}(k)_{22} &= \frac{1}{D(\eta_\ast)}\sqrt{\frac{\eta_\ast}{\eta_{\rm b}}}[J_0(k\eta_\ast)C_Y(\eta_{\rm b})-Y_0(k\eta_\ast)C_J(\eta_{\rm b})],
\end{align}
\end{subequations}
\end{widetext}
where
\begin{align}
C_Y(\eta) &= \frac{1}{2}Y_0(k\eta)+\eta\frac{d}{d\eta}Y_0(k\eta),
\\[0pt]
C_J(\eta) &= \frac{1}{2}J_0(k\eta)+\eta\frac{d}{d\eta}J_0(k\eta),
\\[7pt]
D(\eta) &= C_Y(\eta)J_0(k\eta)-C_J(\eta)Y_0(k\eta).
\end{align}

Similarly in the left stage ($\eta<0$), the general solutions are
\begin{align}
\tilde{u}^{\rm L}_k(\eta) = A^{\rm L}\sqrt{-\eta}J_0(-k\eta) + B^{\rm L}\sqrt{-\eta}Y_0(-k\eta),
\end{align}
where the coefficients $A^{\rm L}$ and $B^{\rm L}$ may be determined by the initial conditions. The corresponding transfer functions ${\bf T}^{\rm L}(k) \equiv {\bf T}^{\rm L}_k(\eta_{\rm i},-\eta_\ast)$ are
\begin{widetext}
\begin{subequations}\label{TL}
\begin{align}
T^{\rm L}(k)_{11} &= \frac{1}{D(-\eta_{\rm i})}\sqrt{\frac{\eta_\ast}{-\eta_{\rm i}}}[C_Y(-\eta_{\rm i})J_0(k\eta_\ast)-C_J(-\eta_{\rm i})Y_0(k\eta_\ast)],
\\[2pt]
T^{\rm L}(k)_{12} &= \frac{1}{D(-\eta_{\rm i})}\sqrt{-\eta_{\rm i}\eta_\ast}[J_0(-k\eta_{\rm i})Y_0(k\eta_\ast)-Y_0(-k\eta_{\rm i})J_0(k\eta_\ast)],
\\[3pt]
T^{\rm L}(k)_{21} &= \frac{1}{D(-\eta_{\rm i})}\frac{1}{\sqrt{-\eta_{\rm i}\eta_\ast}}[C_Y(-\eta_{\rm i})C_J(\eta_\ast)-C_J(-\eta_{\rm i})C_Y(\eta_\ast)],
\\[2pt]
T^{\rm L}(k)_{22} &= \frac{1}{D(-\eta_{\rm i})}\sqrt{\frac{-\eta_{\rm i}}{\eta_\ast}}[J_0(-k\eta_{\rm i})C_Y(\eta_\ast)-Y_0(-k\eta_{\rm i})C_J(\eta_\ast)].
\end{align}
\end{subequations}
\end{widetext}

\begin{figure*}[t!]
	\centering
	\includegraphics[width=0.96\textwidth]{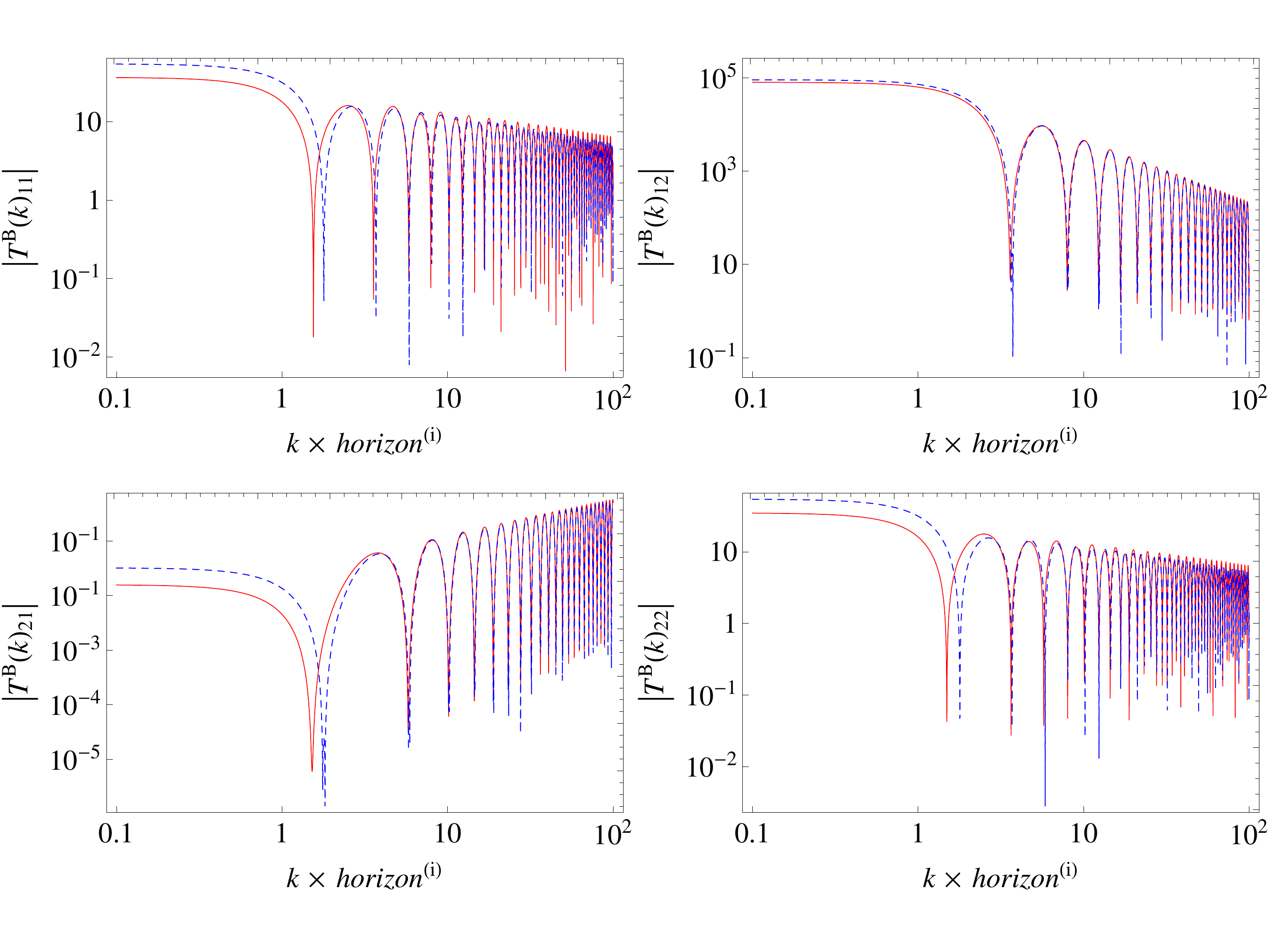}
	\caption{\label{fig03}The numerical (blue-dashed) and analytical (red-solid) results for the transfer functions of GWs in the quantum bounce epoch with Chaotic potential. We have applied the field-free approximation for $m_{\rm eff}$.}
\end{figure*}

Finally according to Eq.~\eqref{eq24}, the overall transfer functions of the quantum bounce epoch can be obtained by combining Eqs.~\eqref{eq52}, \eqref{TR}, and \eqref{TL}:
\begin{align}\label{eq61}
{\bf T}^{\rm B}(k) &\equiv {\bf T}^{\rm B}_k(\eta_{\rm i},\eta_{\rm e}) = {\bf T}^{\rm R}(k)  {\bf T}^{\rm M}(k)  {\bf T}^{\rm L}(k).
\end{align}
Figure \ref{fig03} shows the results. The `$horizon^{\rm (i)}$' in the horizontal axis is the horizon size of causal contact at the beginning of inflation. 
As expected, the field-free approximation works well at large $k$ (small scales). 
To take the advantage of this fact, we can consider 
only the cosmological models where the `$horizon^{\rm (i)}$' is larger than the horizon today so that all the observable GWs today will have wavelengths much smaller than the horizon size at early times and thus eligible for taking the large-$k$ limit. 
In such cases, we not only avoid the inaccuracy shown in Figure \ref{fig03} but also get better physical insight for the evolution of GWs.

The Bessel functions of order zero and their derivatives under the large-$k$ limit are \cite{Abramowitz1972}
\begin{align}
\lim_{k\rightarrow\infty}J_0(k\eta) &= \lim_{k\rightarrow\infty}\frac{d}{kd\eta}Y_0(k\eta) = \sqrt{\frac{2}{\pi k\eta}}\cos\left(k\eta-\frac{\pi}{4}\right),
\\[4pt]
\lim_{k\rightarrow\infty}Y_0(k\eta) &= \lim_{k\rightarrow\infty}\frac{-d}{kd\eta}J_0(k\eta) = \sqrt{\frac{2}{\pi k\eta}}\sin\left(k\eta-\frac{\pi}{4}\right).
\end{align}
Therefore the transfer functions in the right and left stages under the large-$k$ limit are
\begin{align}\label{eq64}
\lim_{k\rightarrow\infty}T^{\rm R}(k) &=
\begin{bmatrix}
  \cos(k\Delta\eta_{\rm R}) & \frac{1}{k}\sin(k\Delta\eta_{\rm R}) \\[3pt]
  -k\sin(k\Delta\eta_{\rm R}) & \cos(k\Delta\eta_{\rm R})
\end{bmatrix},
\\[5pt] \label{eq65}
\lim_{k\rightarrow\infty}T^{\rm L}(k) &=
\begin{bmatrix}
  \cos(k\Delta\eta_{\rm L}) & -\frac{1}{k}\sin(k\Delta\eta_{\rm L}) \\[3pt]
  k\sin(k\Delta\eta_{\rm L}) & \cos(k\Delta\eta_{\rm L})
\end{bmatrix},
\end{align}
where $\Delta\eta_{\rm R}=\eta_{\rm b}-\eta_\ast$ and $\Delta\eta_{\rm L}=-\eta_\ast-\eta_{\rm i}$.

One interesting feature in this result is that the forms of the transfer functions in the left and right stages have a time reversal symmetry. It means that the field-free approximation leads to a time-symmetric quantum bounce effect. This is actually not surprising because according to Section \ref{sec02} the cosmological background should be time-symmetric with respect to the quantum bounce if there is no scalar field.

Under the large-$k$ limit, GWs behave just like plane waves during the entire quantum bounce epoch because the effective mass is negligible. In such case the transfer functions should possess the simple form for plane waves:
\begin{align}\label{eq66}
\lim_{k\rightarrow\infty}T^{\rm B}(k) =
\begin{bmatrix}
  \cos(k\Delta\eta_{\rm B}) & \frac{1}{k}\sin(k\Delta\eta_{\rm B}) \\[3pt]
  -k\sin(k\Delta\eta_{\rm B}) & \cos(k\Delta\eta_{\rm B})
\end{bmatrix},
\end{align}
where $\Delta\eta_{\rm B}=\eta_{\rm b}-\eta_{\rm i}=\Delta\eta_{\rm L}+\Delta\eta_{\rm M}+\Delta\eta_{\rm R}$. 
Indeed this is exactly the same as the result that we obtain from our Eq.~\eqref{eq61} with Eqs.~\eqref{eq52}, \eqref{TR}, and \eqref{TL} under the large-$k$ limit ($k\rightarrow\infty$).

% -----------------------------------------------------------------------
\section{Discussion}\label{sec06}
\subsection{Consistency with the Bogoliubov transformation}\label{sec6a}

We first lay out the framework of the Bogoliubov transformation in literature \cite{Mielczarek2010} and then link it to our new approach of transfer functions for cross checking.
In their approximation for the effective mass, there are three evolutionary stages and hence we need three transition matrices for transforming the creation and annihilation operators. In our notations, they are
\begin{align}
{\bf M}^{\rm V}_k(\eta) &=
\begin{bmatrix}
  \frac{1}{\sqrt{2k}}e^{-ik\eta} & \frac{1}{\sqrt{2k}}e^{ik\eta} \\[4pt]
  \frac{-ik}{\sqrt{2k}}e^{-ik\eta} & \frac{ik}{\sqrt{2k}}e^{ik\eta}
\end{bmatrix},
\\[5pt]
{\bf M}^{\rm B}_k(\eta) &=
\begin{bmatrix}
  \frac{1}{\sqrt{2k_\ast}}e^{-ik_\ast\eta} & \frac{1}{\sqrt{2k_\ast}}e^{ik_\ast\eta}
  \\[4pt]
  \frac{-ik_\ast}{\sqrt{2k_\ast}}e^{-ik_\ast\eta} & \frac{ik_\ast}{\sqrt{2k_\ast}}e^{ik_\ast\eta}
\end{bmatrix},
\\[5pt]
{\bf M}^{\rm I}_k(\eta) &=
\begin{bmatrix}
  g_k(\eta) & g^\ast_k(\eta) \\[3pt]
  \frac{d}{d\eta}g_k(\eta) & \frac{d}{d\eta}g^\ast_k(\eta)
\end{bmatrix},
\end{align}
where the mode function $g_k(\eta)$ is
\begin{align}
g_k(\eta) = -\sqrt{\frac{-\pi\eta_{-\rm e}}{4}}H_{3/2}^{(1)}(k\eta_{-\rm e}),
\end{align}
which satisfies Eq.~\eqref{eq38} with
\begin{align}
\hat{u}^{\rm I}_k(\eta) = g_k(\eta) \hat{b}_k + g_k^\ast(\eta) \hat{b}_{-k}^\dagger.
\end{align}
Here $H^{(1)}$ is the Hankel function of the first kind, $\eta_{-\rm e}=\eta-\eta_{\rm e}$, the superscripts `V' and `B' stand for the Minkowski vacuum and the quantum bounce respectively. The corresponding mode function $f_k$ satisfying the plane wave equation is given by \cite{Mielczarek2010}
\begin{align}
\hat{u}^{\rm V}_k(\eta) = f_k(\eta) \hat{a}_k + f_k^\ast(\eta) \hat{a}_{-k}^\dagger.
\end{align}
Conventionally $\hat{a}_{-k}^\dagger$ and $\hat{a}_{k}$ are the creation and annihilation operators for the Minkowski vacuum. Similarly $\hat{b}_{-k}^\dagger$ and $\hat{b}_{k}$ are the creation and annihilation operators for inflation. The mode functions $f_k$ and $g_k$ must satisfy the Wronskian condition \cite{Mielczarek2010}
\begin{align}
W(f_k,f_k^\ast) = W(g_k,g_k^\ast) = i.
\end{align}

The Bogoliubov transformation is therefore
\begin{align}
\begin{bmatrix}
  \hat{b}_{k} \\[3pt] \hat{b}_{-k}^\dagger
\end{bmatrix} =
\begin{bmatrix}
  \alpha_k & \beta_k^\ast \\[3pt]
  \beta_k & \alpha_k^\ast
\end{bmatrix}
\begin{bmatrix}
  \hat{a}_{k} \\[3pt] \hat{a}_{-k}^\dagger
\end{bmatrix}.
\end{align}
The coefficients $\alpha_k$ and $\beta_k$ can then be determined as 
\begin{align}
\begin{bmatrix}
  \alpha_k \\[2pt] \beta_k
\end{bmatrix} =
{\bf M}^{\rm I}_k(\eta_{\rm b})^{-1}{\bf M}^{\rm B}_k(\eta_{\rm b}){\bf M}^{\rm B}_k(\eta_{\rm i})^{-1}{\bf M}^{\rm V}_k(\eta_{\rm i})
\begin{bmatrix}
  1 \\[2pt] 0
\end{bmatrix}.
\end{align}
Finally, the GW power spectrum can be obtained by
\begin{align}
P_{\rm T}(k) = |\alpha_k - \beta_k|^2P_{\rm T}^{\rm std}(k).
\end{align}

Two interesting features can be revealed here. First, the initial GWs generated by the quantum fluctuations in the Minkowski vacuum can now be derived as
\begin{align}
{\bf M}^{\rm V}_k(\eta_{\rm i})
\begin{bmatrix}
  1 \\[2pt] 0
\end{bmatrix} =
\begin{bmatrix}
  \frac{1}{\sqrt{2k}} \\[4pt] \frac{-ik}{\sqrt{2k}}
\end{bmatrix}
e^{-ik\eta_{\rm i}} = \tilde{U}^{\rm V}_k(\eta_{\rm i}).
\end{align}
Secondly, the transfer functions of the quantum bounce epoch with the approximation $m_{\rm eff}^2=k_0^2$ (for $\eta_{\rm i} < \eta < \eta_{\rm b}$) \cite{Mielczarek2010} are now given as
\begin{align}\label{eq3.73}
{\bf M}^{\rm B}_k(\eta_{\rm b}){\bf M}^{\rm B}_k(\eta_{\rm i})^{-1} &= 
\begin{bmatrix}
  \cos(k_\ast\Delta\eta_{\rm B}) & \frac{1}{k_\ast}\sin(k_\ast\Delta\eta_{\rm B}) \\[3pt]
  -k_\ast\sin(k_\ast\Delta\eta_{\rm B}) & \cos(k_\ast\Delta\eta_{\rm B})
\end{bmatrix} \nonumber \\
&\equiv T^{\rm B,old}(k).
\end{align}
The superscript `old' stands for the results based on the approximation in literature \cite{Mielczarek2010}. We shall discuss this in more details in Section \ref{ch3.5.2}. 

\begin{figure}[t!]
	\centering
	\includegraphics[width=0.48\textwidth]{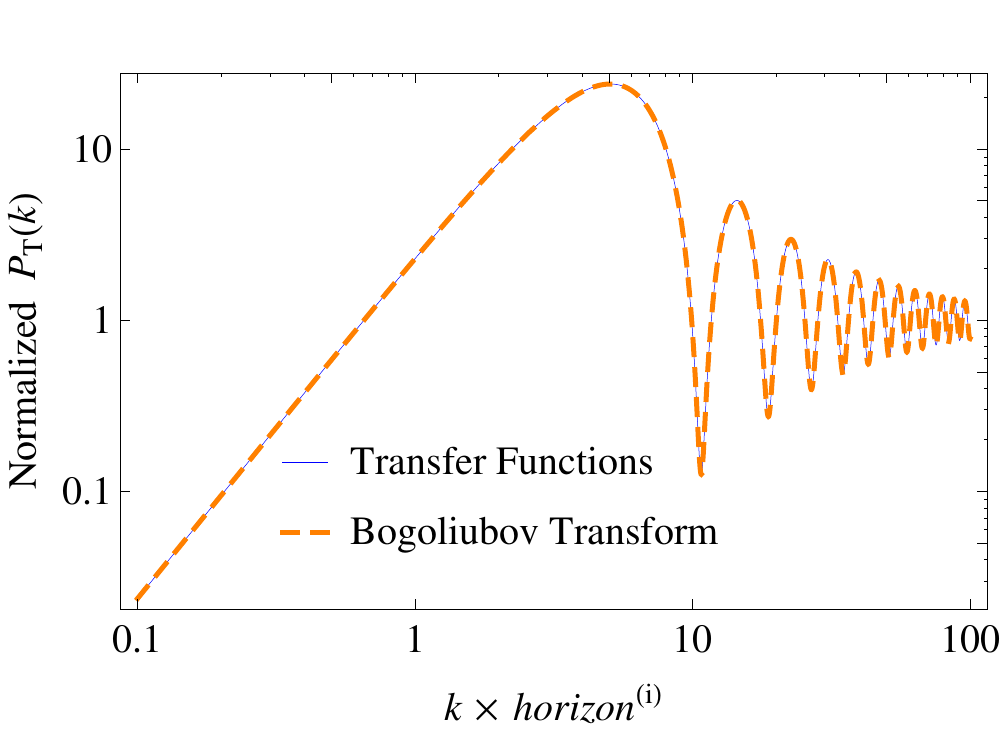}
	\caption{\label{fig04}The GW power spectra generated by the quantum fluctuations in the Minkowski vacuum before the quantum bounce epoch. The results based on the Bogoliubov transformation (dashed) and on our approach of transfer functions (solid) are totally consistent.
	}
\end{figure}

Figure \ref{fig04} compares the GW power spectra based on the Bogoliubov transformation (dashed curve) and on our approach of transfer functions (solid curve). It is obvious that they are totally consistent with each other.

\subsection{Resolving the IR suppression problem}\label{ch3.5.2}

In Figure \ref{fig04}, there is a clear suppression in the normalized GW power spectrum towards long wavelengths (small $k$). This is called the IR suppression first pointed out by Ref.~\cite{Mielczarek2010}. In this section, we shall employ our transfer functions to show that the IR suppression is not of a physical origin but a consequence of improper normalization.

\begin{figure}[t!]
	\centering
	\includegraphics[width=0.48\textwidth]{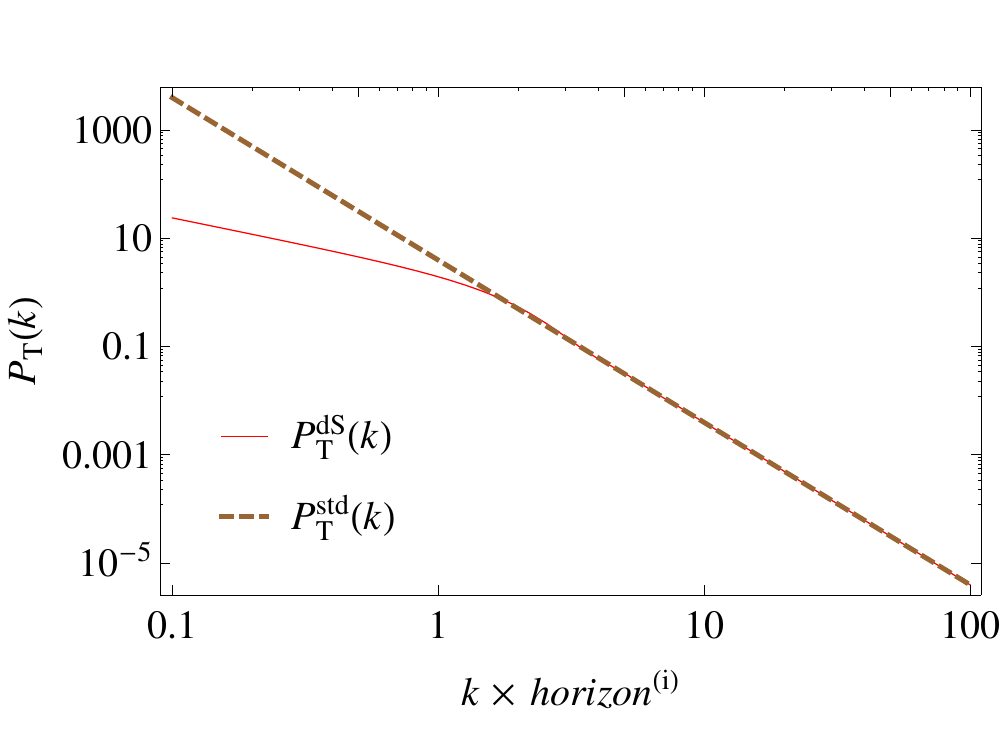}
	\caption{\label{fig05}The primordial GW power spectra in the standard inflationary cosmology (solid) as compared with the result based on sub-horizon approximation (dashed).}
\end{figure}

The problem comes from the fact that the GW power spectrum in Figure \ref{fig04} is normalized to Eq.~\eqref{eq16}, which is valid only for the sub-horizon modes (large $k$). To reveal the super-horizon behavior of the primordial perturbations, we take the small-$k$ limit for Eq.~\eqref{eq44}:
\begin{align}\label{eq3.75}
\lim_{k\rightarrow 0} T^{\rm I}(k) = \frac{1}{3}
\begin{bmatrix}
	2a_{\rm e}H_{\rm dS}\Delta\eta_{\rm I} & -a_{\rm e}H_{\rm dS}\Delta\eta_{\rm I}^2 \\
	-2a_{\rm e}^2H_{\rm dS}^2\Delta\eta_{\rm I} & a_{\rm e}^2H_{\rm dS}^2\Delta\eta_{\rm I}^2
\end{bmatrix}.
\end{align}
Then the GW power spectrum generated by the quantum fluctuations in de Sitter space is
\begin{align}\label{eq3.76}
\lim_{k\rightarrow 0} P_{\rm T}^{\rm dS}(k)= \frac{32\pi H_{\rm dS}^2\Delta\eta_{\rm I}^2}{9k} \propto k^{-1}.
\end{align}
Eqs.~\eqref{PT-sub} and \eqref{eq3.76} together conclude that
$P_{\rm T}^{\rm dS} \propto k^{-3}$ on sub-horizon scales (large $k$) and
$P_{\rm T}^{\rm dS} \propto k^{-1}$ on super-horizon scales (small $k$).
These behaviors can be seen in the red solid curve in Figure \ref{fig05}. The transition of the two behaviors occurs at $k\times horizon^{\rm (i)}\sim 1$.
On super-horizon scales, this exact solution deviates significantly from the pure sub-horizon approximation (dashed curve).
In other words, taking the $k^{-3}$ dependence throughout as the primordial form to normalize the entire power spectrum would cause the deficit in the power at small $k$,
and thus the IR suppression problem.

\begin{figure}[t!]
	\centering
	\includegraphics[width=0.48\textwidth]{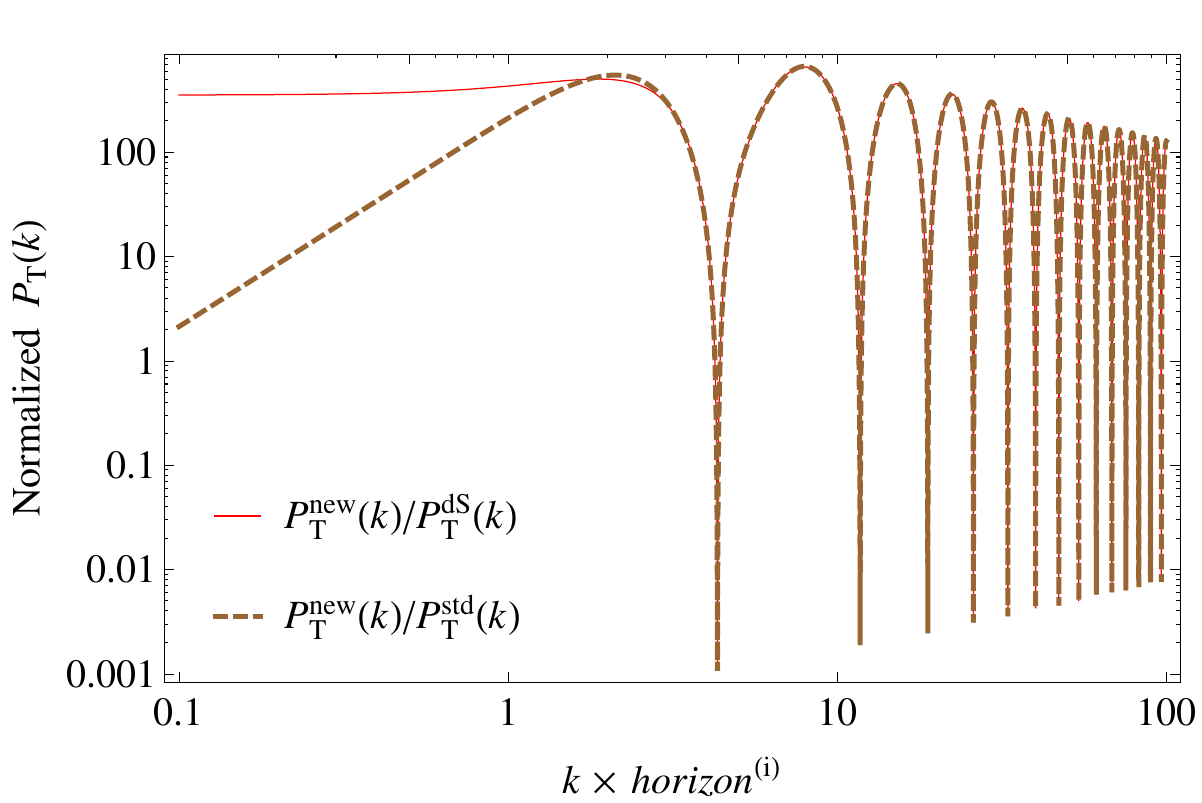}
	\caption{\label{fig06}The GW power spectrum normalized to its primordial power spectrum in a realistic cosmological model.
	The brown dashed curve is normalized to $P_{\rm T}^{\rm std}(k)\propto k^{-3}$ (sub-horizon approximation) while the red solid curve is normalized to the correct $P_{\rm T}^{\rm dS}(k)$.
	}
\end{figure}

Figure \ref{fig06} shows the GW power spectrum normalized to its primordial power spectrum in a realistic cosmological background. 
The brown dashed curve is normalized to $P_{\rm T}^{\rm std}(k)\propto k^{-3}$ (sub-horizon approximation) so the expected IR suppression problem at small $k$ is clear. On the other hand, the red solid curve is normalized to the correct $P_{\rm T}^{\rm dS}(k)$ based on our formalism of transfer functions, so the IR suppression problem disappears.
The correctly normalized GW power spectrum remains constant on the super-horizon scales while oscillating on sub-horizon scales. It means that the GWs originally frozen outside the horizon will start oscillating after the horizon entry.

\subsection{Improvement from the field-free approximation}

We now demonstrate how much our formalism could improve on the accuracy in evolving the GWs, as compared with the literature \cite{Mielczarek2010}.
In addition to the transparency of the transfer function formalism in evolving the GWs, another major contribution of our work is the field-free approximation, which provides a better estimate of the effective mass so as to improve the accuracy in evolving the GWs during the quantum bounce epoch.

\begin{figure}[t!]
	\centering
	\includegraphics[width=0.48\textwidth]{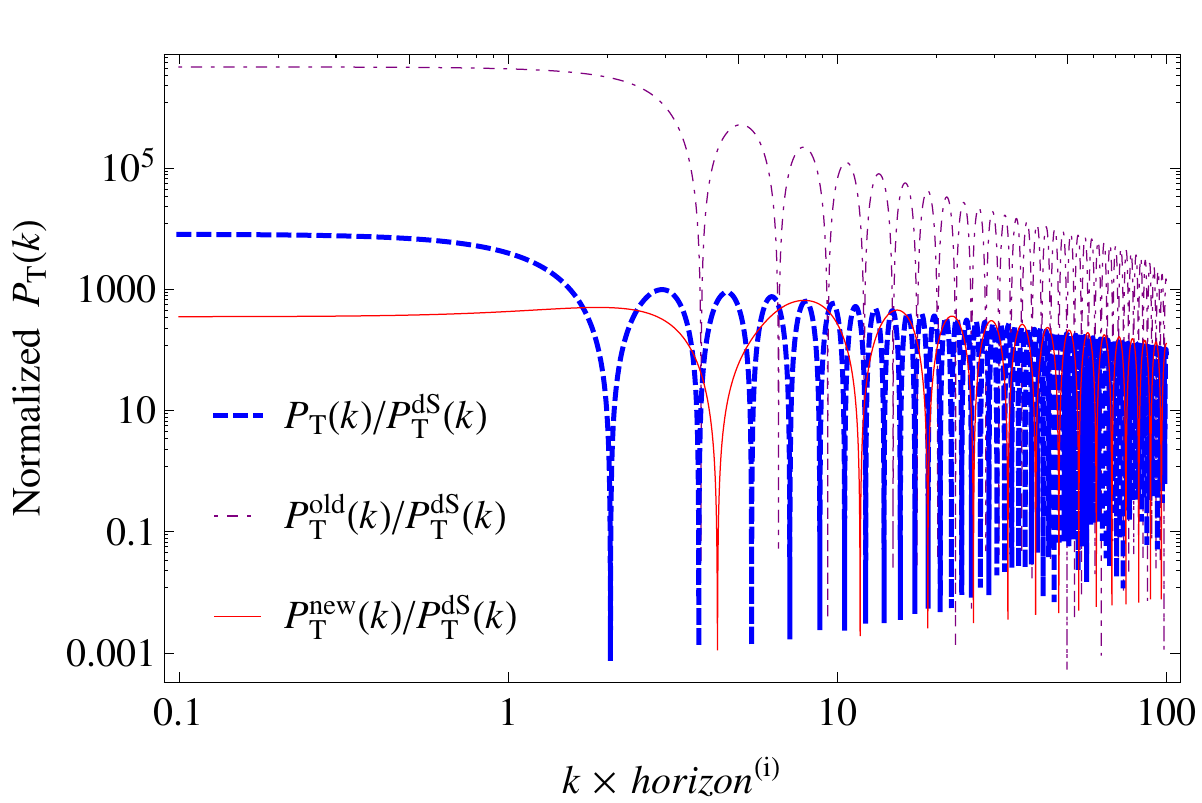}
	\caption{\label{fig07}The predicted GW power spectra based on the `old' (dot-dashed) and `new' (solid) approaches as compared with the numerical result (dashed), all normalized to the same primordial $P_{\rm T}^{\rm dS}(k)$.}
\end{figure}

Ref.~\cite{Mielczarek2010} assumes the effective mass to be a constant $k_0$ throughout the quantum bounce epoch. This leads to Eq.~\eqref{eq3.73} in our formalism using transfer functions. In turn we could obtain the GW power spectrum, which is shown in Figure \ref{fig07} as the dot-dashed curve, labeled with `old'. 
On the other hand, our field-free approximation, which divide and estimate the evolutionary behavior of the effective mass into three stages (see Eq.~\eqref{field-free} and Figure \ref{fig01}), leads to the solid curve in Figure \ref{fig07}, labeled with `new'. 
These two results are compared to the numerical solution (dashed curve).
It is evident that our `field-free' approximation leads to a dramatical improvement, while the approximation in Ref.~\cite{Mielczarek2010} results in an overestimation by an order of about two. 
Although our result may be further improved on super-horizon scales and on the phase accuracy, these issues are not critical as the causal horizon today is required to be smaller than that at the beginning of inflation so that the observable scales today should be at the large-$k$ end of the diagram, where our approximation delivers a good result.

% -----------------------------------------------------------------------
\section{Models with a parent universe}\label{sec08}
\subsection{The bouncing scenarios}\label{sec08a}

Previously in Eq.~\eqref{eq14}, we have introduced $\theta_{\rm B}$ to discribe the time-symmetry of cosmological background. However $\theta_{\rm B}$ doesn't work well for a potential which is not always positive because the sine term goes to imaginary when the potential goes below zero at the quantum bounce. We therefore directly use, instead of $\theta_{\rm B}$, the value of scalar field at the quantum bounce $\phi_{\rm B}$ to quantify the time-symmetry of cosmological background. We now have, for example for Chaotic potential, two parameters $m_\phi$ and $\phi_{\rm B}$ that determine the evolution of cosmological background and thus determine the transfer functions.

\begin{figure}[t!]
	\centering
	\includegraphics[width=0.48\textwidth]{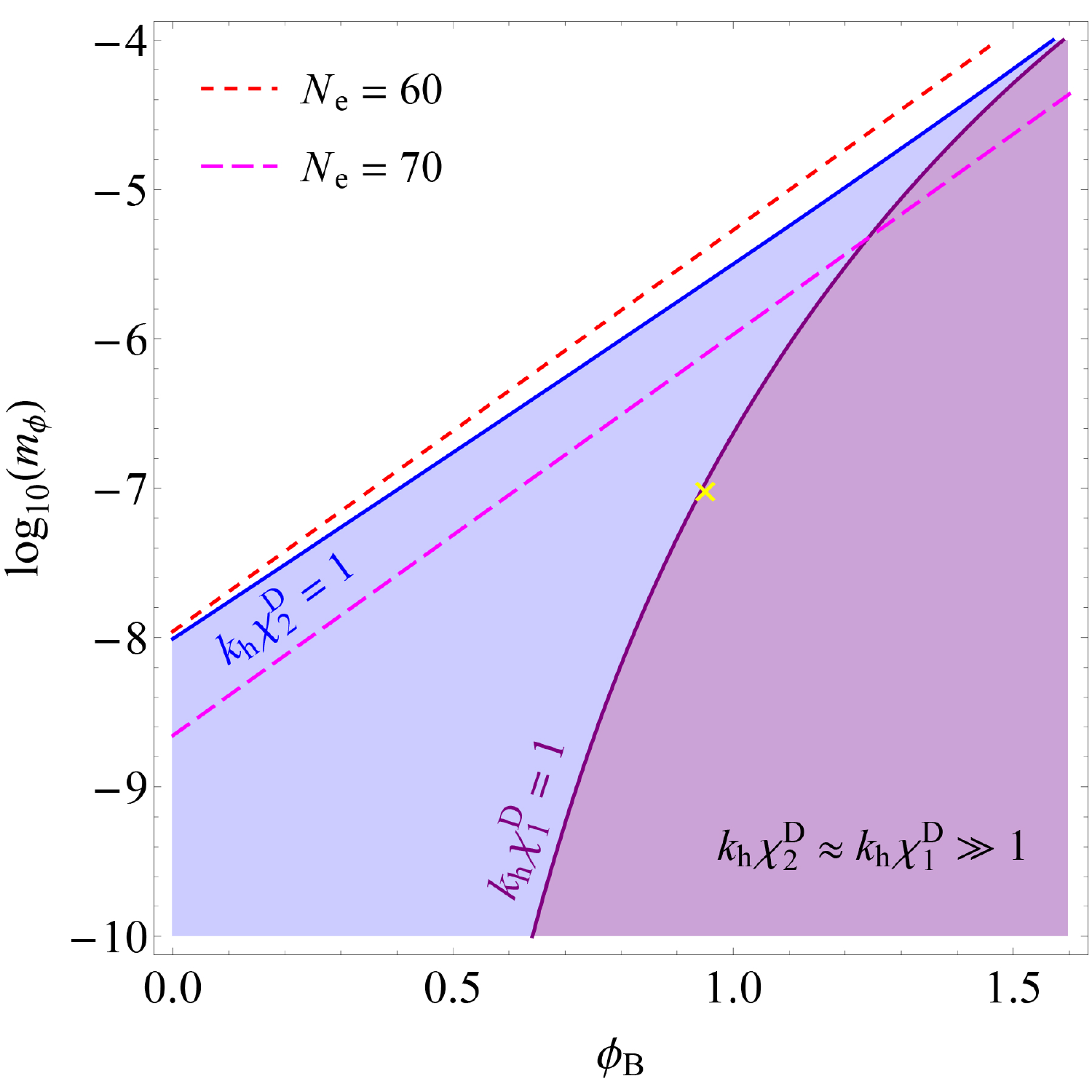}
	\caption{\label{fig08}The HABS (purple), IABS (blue), and NSBS (blank) regions in the parameter space $(\phi_{\rm B},m_\phi)$. The two dashed lines are drawn for $N_e=60$ and $70$, respectively.}
\end{figure}

In this section we consider the models with a parent universe. In other words, there is deflation comes before the quantum bounce epoch instead of the Minkowski vacuum that we considered since Section \ref{sec05}. The quantum bounce epoch therefore starts at the end of deflation $\eta=\eta_{\rm e}^{\rm D}$. There are in terms seven resulted quantities that affect the behaviors of transfer functions:
$\{ \chi_2^{\rm D}, \chi_1^{\rm D}, \eta_{\rm b}^{\rm D}, \eta_\ast, \eta_{\rm b}, \chi_2^{\rm I}, \chi_1^{\rm I} \}$,
where $\chi_1^{\rm D}\equiv-(a_{\rm b}^{\rm D}H_{\rm dS}^{\rm D})^{-1}$, $\chi_2^{\rm I}\equiv(a_{\rm e}H_{\rm dS})^{-1}$.
We can then classify the bouncing scenarios for $\phi_{\rm B}>0$ into three types: highly asymmetric bouncing scenario (HABS), intermediately asymmetric bouncing scenario (IABS), and nearly symmetric bouncing scenario (NSBS), by the following conditions:
\begin{align*}
	k_{\rm h}\chi_2^{\rm D} >1 \mbox{~~and~~} k_{\rm h}\chi_1^{\rm D} >1 & \mbox{~~for HABS}, \\
	k_{\rm h}\chi_2^{\rm D} >1 \mbox{~~and~~} k_{\rm h}\chi_1^{\rm D} <1 & \mbox{~~for IABS}, \\
	k_{\rm h}\chi_2^{\rm D} <1 \mbox{~~and~~} k_{\rm h}\chi_1^{\rm D} <1 & \mbox{~~for NSBS},
\end{align*}
where $k_{\rm h}$ is the comoving wavenumber of the Hubble horizon today. Figure \ref{fig08} shows the regions of these three scenarios in the two-dimensional parameter space $(\phi_{\rm B},m_\phi)$ where the solid curves underline the boundaries. It is clear that most of the scenarios with number of e-foldings for inflation $N_e$ greater than $60$ are HABS's and IABS's.
In these two scenarios, inflation is strong enough that $k_{\rm h}\chi_1^{\rm I}$ is much greater than one and $k_{\rm h}\chi_2^{\rm I}$ is much smaller than one so that $T^{\rm I}(k)$ must follow Eq.~\eqref{eq43}. On the other hand, the quantum bounce occurs at a relatively small scale so $T^{\rm B}(k)$ satisfies the large-$k$ limit and thus follows Eq.~\eqref{eq66}. The evolutions of pre-existing GWs are therefore simply determined only by deflation. In the following section, we combine all of these conditions and take corresponding limits on $T^{\rm D}(k)$ for both HABS and IABS to study the behaviors of pre-existing GWs.

\subsection{Signals from the parent universe}\label{sec08a}

First of all we consider the HABS. In this kind of cosmological background, deflation is small that GWs propagate like plane waves in this era. The transfer functions for deflationary epoch are therefore reduced to
\begin{align}\label{eq3.77}
	T^{\rm D}(k) = \left[
	\begin{matrix}
		\cos(k\Delta\eta_{\rm D})  & \frac{1}{k}\sin(k\Delta\eta_{\rm D}) \\[3pt]
		-k\sin(k\Delta\eta_{\rm D}) & \cos(k\Delta\eta_{\rm D})
	\end{matrix}
	\right] 
\end{align}
and the resulting matrix of overall transfer functions is
\begin{align}\label{eq3.78}
	{\bf T}(k) = 
	\begin{bmatrix}
	  -\frac{a_{\rm e}H_{\rm dS}}{k}\sin(k\Delta\eta_{\rm all}) 		&
	   \frac{a_{\rm e}H_{\rm dS}}{k^2}\cos(k\Delta\eta_{\rm all})		\\[3pt]
	   \frac{a_{\rm e}^2H_{\rm dS}^2}{k}\sin(k\Delta\eta_{\rm all}) 	&
	  -\frac{a_{\rm e}^2H_{\rm dS}^2}{k^2}\cos(k\Delta\eta_{\rm all})
	\end{bmatrix},
\end{align}
where $\Delta\eta_{\rm all}=\Delta\eta_{\rm D}+\Delta\eta_{\rm B}+\Delta\eta_{\rm I}$. We find that Eq.~\eqref{eq3.78} takes exactly the same forms as ${\bf T}^{\rm I}(k)$ but with a different phase. With initial conditions $\tilde{u}_k^{\rm (i)}$ and its derivative $-ik\tilde{u}_k^{\rm (i)}$, the power spectrum is given by
\begin{align}
	P_{\rm T}(k) = \frac{16\pi H_{\rm dS}^2}{k^2}|\tilde{u}_k^{\rm (i)}|^2 \propto k^{-2}|\tilde{u}_k^{\rm (i)}|^2.
\end{align}
This result indicates that, in the HABS, deflation is too weak to generate distinguishable feachers on GWs.

We then head into the IABS where deflation become much stronger such that $k_{\rm h}\chi_1^{\rm D}$ goes far below one. The corresponding transfer functions for deflationary epoch are
\begin{align}
	{\bf T}^{\rm D}&(k) =
	\begin{bmatrix}
	  -\frac{\left( a_{\rm b}^{\rm D}H_{\rm dS}^{\rm D}\right)^2 }{k^2}\cos(k\Delta\eta_{\rm D})	&
	  -\frac{a_{\rm b}^{\rm D}H_{\rm dS}^{\rm D}}{k^2}\cos(k\Delta\eta_{\rm D}) 					\\[3pt]
	   \frac{\left( a_{\rm b}^{\rm D}H_{\rm dS}^{\rm D}\right)^2 }{k}\sin(k\Delta\eta_{\rm D})		&
	   \frac{a_{\rm b}^{\rm D}H_{\rm dS}^{\rm D}}{k}\sin(k\Delta\eta_{\rm D})
	\end{bmatrix}
\end{align}
and the matrix of overall transfer functions is therefore
\begin{align}\label{eq3.78}
	{\bf T}(k) = 
	\begin{bmatrix}
	   \frac{\left( a_{\rm b}^{\rm D}H_{\rm dS}^{\rm D}\right)^2 }{k^2}\frac{a_{\rm e}H_{\rm dS}}{k} 		&
	   \frac{a_{\rm b}^{\rm D}H_{\rm dS}^{\rm D}}{k^2}\frac{a_{\rm e}H_{\rm dS}}{k}							\\[3pt]
	  -\frac{\left( a_{\rm b}^{\rm D}H_{\rm dS}^{\rm D}\right)^2 }{k^2}\frac{a_{\rm e}^2H_{\rm dS}^2}{k} 	&
	  -\frac{a_{\rm b}^{\rm D}H_{\rm dS}^{\rm D}}{k^2}\frac{a_{\rm e}^2H_{\rm dS}^2}{k}
	\end{bmatrix}
	\sin(k\Delta\eta_{\rm all}^\ast),
\end{align}
where $\Delta\eta_{\rm all}^\ast=\Delta\eta_{\rm D}+\Delta\eta_{\rm B}-\Delta\eta_{\rm I}$.
It's obvious that the evolution of GWs in IABS is significantly different from HABS especially on the $k$ dependence. By applying the same initial conditions $\tilde{u}_k^{\rm (i)}$ and its derivative $-ik\tilde{u}_k^{\rm (i)}$, we can see that the power spectrum
\begin{align}
	P_{\rm T}(k) &=
		\left( \frac{a_{\rm b}^{\rm D}H_{\rm dS}^{\rm D}}{k} \right)^4 \sin^2(k\Delta\eta_{\rm all}^\ast)
		\frac{16\pi H_{\rm dS}^2}{k^2}|\tilde{u}_k^{\rm (i)}|^2 \nonumber \\
		&\propto k^{-6}\sin^2(k\Delta\eta_{\rm all}^\ast)|\tilde{u}_k^{\rm (i)}|^2
\end{align}
tilts much more toward large $k$. In other words, deflation suppresses small scale waves and enhances large scale waves in the meantime. It means that in our context, we are more possible to observe the pre-existing GWs on large scales rather than small scales.

\begin{figure}[t!]
	\centering
	\includegraphics[width=0.48\textwidth]{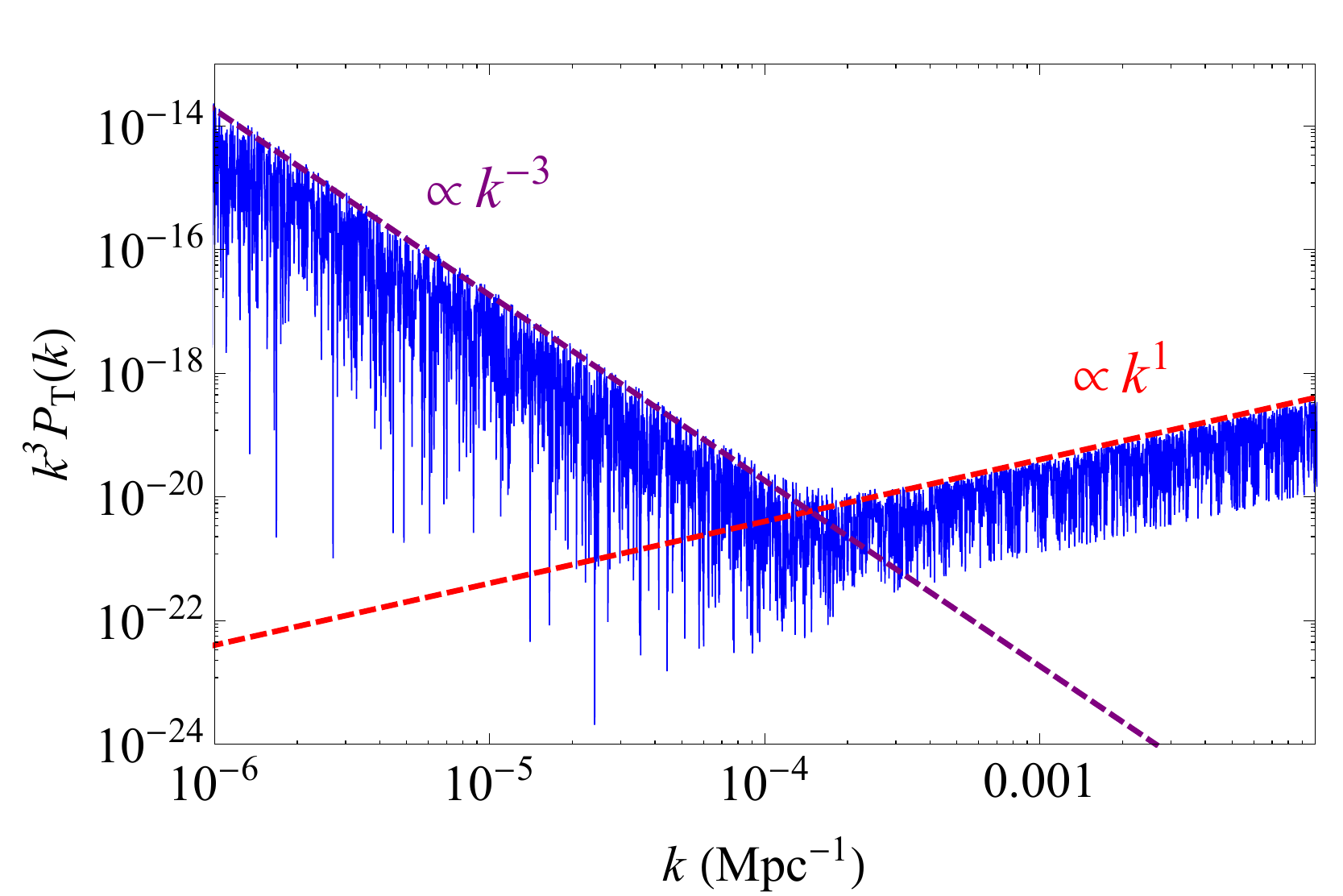}
	\caption{\label{fig09}The scale-invariant power spectrum of pre-existing GWs generated by a constant $\tilde{u}_k^{\rm (i)}$ in the parent universe.}
\end{figure}

In order to discuss and compare these two scenarios at once, we consider a case with $(\phi_{\rm B},m_\phi)=(0.95,10^{-7})$ marked by the yellow cross in Figure \ref{fig08}, which appears near the boundary of HABS and IABS, and see how $P_{\rm T}(k)$ generated by a constant $\tilde{u}_k^{\rm (i)}$ behaves. The result is shown in Figure \ref{fig09} with $k_{\rm h}\approx 0.0002~\rm{Mpc}^{-1}$.
The sub-horizon waves which correspond to the results of $k_{\rm h}\chi_1^{\rm D}>1$ behave like in HABS, where $k^3P_{\rm T}(k)\propto k$. In contrast, the super-horizon waves which correspond to the results of $k_{\rm h}\chi_1^{\rm D}<1$ behave like in IABS, where $k^3P_{\rm T}(k)\propto k^{-3}$. We conclude that the effects from deflation is invisible if the bouncing scenario is highly asymmetric, and expected on large scale if the bouncing scenario is intermediately asymmetric. We note that the number of e-foldings is too small in the NSBS which we do not discuss here in this paper.

Our transfer functions is mathematically and physically transparent that if we generally consider an arbitrary source of pre-existing GWs in the parent universe, say $\tilde{u}_k^{\rm (i)}\propto k^n$, the resulted power spectrums is simply computed as $k^3P_{\rm T}(k)\propto k^{2n+1}$ for HABS and $k^3P_{\rm T}(k)\propto k^{2n-3}$ for IABS. This helps us to study other possible kinds of pre-existing GWs such as those originated from well-known astronomical systems. Indeed we have used the transfer functions to obtain the observational features resulting from the GWs generated by stellar binaries in the parent universe in our another work which is going to be published soon.

% -----------------------------------------------------------------------
\section{Conclusion}\label{sec07}

In this paper, we propose a new formalism that employs the transfer functions to evolve the GWs. For the first time in literature this enables us to study the GWs even before the quantum bounce.
We offer the analytical forms of the transfer functions for the quantum bounce epoch, the inflationary epoch, and the possible deflationary epoch. We also provide a complete recipe for obtaining accurate numerical results and utilize it for our discussions and to verify our results.
In particular our transfer functions deliver accurate evolution for the primordial GW power spectrum in the standard inflationary cosmology,
and our field-free approximation for the effective mass in the quantum bounce epoch dramatically improves the accuracy in the predicted GW power spectrum.

We conclude our work by summarizing three essential advantages of the transfer function formalism in evolving the GWs and obtaining their power spectrum.

\begin{enumerate}[(i)]
\item Simplicity: The transfer functions can bring the GWs from the initial state through various epochs to the final state in a simple but accurate manner. It takes the advantage of the linear algebra.

\item Intuitiveness: The transfer functions enable us to deal with the problems of the quantum mechanics in a classical way. We have verified that they are a complete alternative to the Bogoliubov transformations so that we do not need to calculate the Bogoliubov transformations anymore.
In addition, the $k$-dependence in the GW power spectrum can be transparently revealed and we have take this advantage to resolve the IR suppression problem.

\item Generality: The transfer function formalism is general in dealing with any linear problems. In cosmology, most perturbations of our interest are tiny and have evolved mostly in the linear regime. Thus we expect our formalism to be equally useful under other context with linearity.
\end{enumerate}

% -----------------------------------------------------------------------
\appendix
\section{A toy example for deriving the transfer functions of equation of motion}\label{app03}

For demonstration, we consider a simple harmonic oscillator $x=x(t)$ which follows the equation of motion
\begin{align}
\ddot{x} + \omega^2 x = 0.
\end{align}
The solution is well-known as
\begin{align}
x(t) = A\cos\omega t + B\sin\omega t,
\end{align}
where $A$ and $B$ are arbitrary coefficients, which may be determined by the initial conditions at $t_{\rm i}$ through
\begin{align}
\left\{
\begin{array}{l}
	x(t_{\rm i}) = A\cos\omega t_{\rm i} + B\sin\omega t_{\rm i}, \\
	\dot{x}(t_{\rm i}) = -\omega A\sin\omega t_{\rm i} + \omega B\cos\omega t_{\rm i},
\end{array}\right.
\end{align}
which lead to
\begin{align}
\left\{
\begin{array}{l}
	A = \cos\omega t_{\rm i} x(t_{\rm i}) - \frac{\sin\omega t_{\rm i}}{\omega}\dot{x}(t_{\rm i}), \\
	B = \sin\omega t_{\rm i} x(t_{\rm i}) + \frac{\cos\omega t_{\rm i}}{\omega}\dot{x}(t_{\rm i}).
\end{array}\right.
\end{align}
At the final time $t_{\rm f}$ we then have
\begin{align}
\left\{
\begin{array}{l}
	x(t_{\rm f}) = \cos(\omega\Delta t)x(t_{\rm i}) + \frac{1}{\omega}\sin(\omega\Delta t)\dot{x}(t_{\rm i}), \\
	\dot{x}(t_{\rm f}) = -\omega \sin(\omega\Delta t)x(t_{\rm i}) + \cos(\omega\Delta t)\dot{x}(t_{\rm i}).
\end{array}\right.
\end{align}
Therefore the transfer function matrix for the equation of motion is
\begin{align}
{\bf T}_\omega(t_{\rm i},t_{\rm f}) = 
\begin{bmatrix}
 \cos(\omega\Delta t)  & \frac{1}{\omega}\sin(\omega\Delta t) \\
-\omega\sin(\omega\Delta t) & \cos(\omega\Delta t)
\end{bmatrix},
\end{align}
where $\Delta t\equiv t_{\rm f}-t_{\rm i}$. We note that this is obviously a rotation matrix of $SO(2)$ group which rotates the coordinates in the Cartesian plane counter-clockwise by an angle $\omega\Delta t$ about the origin in Euclidean space.

% -----------------------------------------------------------------------

\acknowledgments
We thank Dr. Dah-Wei Chiou for helpful discussions and information at the very beginning of this project. We also acknowledge the support from the Ministry of Science and Technology, Taiwan (NSC 100-2112-M-002-004-MY3; MOST 103-2628-M-002-006-MY4).

% -----------------------------------------------------------------------

\bibliography{reference_01}

\end{document}